# Design and Evaluation of Mechanisms for a Multicomputer Object Store

Lex Weaver

November 28, 1994

## Abstract

Multicomputers have traditionally been viewed as powerful compute engines. It is from this perspective that they have been applied to various problems in order to achieve significant performance gains. There are many applications for which this compute intensive approach is only a partial solution. CAD, virtual reality, simulation, document management and analysis all require timely access to large amounts of data. This thesis investigates the use of the object store paradigm to harness the large distributed memories found on multicomputers. The design, implementation, and evaluation of a distributed object server on the Fujitsu AP1000 is described. The performance of the distributed object server under example applications, mainly physical simulation problems, is used to evaluate solutions to the problems of client space recovery, object migration, and coherence maintenance.

The distributed object server follows the client-server model, allows object replication, and uses binary semaphores as a concurrency control measure. Instrumentation of the server under these applications supports several conclusions: client space recovery should be dynamically controlled by the application, predictively prefetching object replicas yields benefits in restricted circumstances, object migration by storage unit (segment) is not generally suitable where there are many objects per storage unit, and binary semaphores are an expensive concurrency control measure in this environment.

# Acknowledgements

Many thanks to my supervisor, Dr Chris Johnson, for introducing me to the world of parallel computing, and much helpful advice throughout the past year. Thank you also to my parents for financing me through a seemingly interminable degree, to my flatmate Rachel for tolerating the vagaries of a computing honours student, and to the Canberra Workers' Club for their five dollar steak special.

# Contents

# List of Figures

# List of Tables

# Chapter 1

# Introduction

Applications requiring fast access to huge amounts of persistent data are known as *data hungry* applications. This project explores issues relevant to the construction of a multicomputer object store to be utilised as a support mechanism for these applications.

## 1.1 Motivation

Text management and analysis, scientific simulations, virtual reality, and CAD are all examples of data hungry applications. They have an appetite for data that consumes available storage and transfer capacity. As these applications grow in complexity and their data requirements increase, new support mechanisms will be required to achieve acceptable performance.

The current crop of high performance machines have a distributed memory MIMD architecture and are typified by the Fujitsu AP1000 and IBM SP2. With many megabytes of memory per processor, such machines possess total memories measured in gigabytes, easily dwarfing the half-gigabyte available in high-end workstations. In order to support data hungry applications, these large distributed memories need to be harnessed with a support mechanism capable of delivering good performance.

The distributed object store is the paradigm under consideration; it provides an abstraction similar to the heap used in many procedural languages, but is not tightly coupled to the Private Address Space (PAS), linear memory model [29]. It aims to provide support for cooperative, information intensive tasks and is often seen as an integration of database and programming language features [27]. Specifically, it seeks to provide object-oriented access to large amounts of persistent data in a manner which is fast, concurrent and transparent. Objects are units of related data, similar to records or structures in procedural languages. Fast access is a prerequisite for good performance, concurrent access allows the full potential of the multicomputer architecture to be utilised, and transparency seeks to provide a simple programming paradigm. Brown [9] has estimated that up to 30% of code in an application program is concerned with persistence. Making object access transparent with respect to its location, reduces the size of this portion of the code and frees the programmer from the need to understand and implement data transfer and translation methodologies [36].

The following sections of this chapter outline issues relevant to a multicomputer object store and indicate the scope of this investigation.

## 1.2 Issues

There are many issues concomitant with a distributed object store. The broadest are outlined below, and are classified as functional (to do with the logical model presented to the programmer) or implementation (affecting only the manner in which the operations of the logical model are performed). Issues from each class have the potential to impact upon performance.

Many of the issues are discussed in the literature under topics as diverse as communications protocols, databases, and shared memory. Chapter 2 presents a survey of some of the existing literature relevant to the issues within the scope of the project.

### 1.2.1 Functional Issues

The abstract model presented to the programmer must be capable of expressing the required semantics in a clear manner using primitives which are efficiently implemented. Which primitives to provide, and their exact semantics, form several significant questions.

#### object identification

How to identify and address objects is a significant problem. An object store is conceptually infinite. This and the proposed store's distributed nature make the linear memory model unsuitable. Relevant questions include:

- whether to generate identifiers, or allow the user application to provide names

- are identifiers/names unique?

- can identifiers/names be reused?

- if identifiers can be removed from the system's control, when is it safe to delete the associated object from the store?

The next chapter addresses object identity in further detail and considers the advantages of *referential integrity* over *name based association.*

**what operations to provide?**

Simple read and write are obvious choices, but more complex operations such as find and replace may be often used and yield better performance if provided as primitives.

**space recovery in the store**

In order to simulate an infinite store, it is necessary to recover the space occupied by objects which are no longer required. How the identity of these objects is determined, and when they can safely be removed are interesting questions addressed further in the next chapter.

**how to maintain coherence under concurrent access?**

Many models exist for the maintenance of data coherence under concurrent access. *Strict coherence* may be maintained using explicit object locking of various forms, or the higher level transaction model implemented using optimistic scheduling or a locking mechanism. *Loose coherence* mechanisms avoid the overheads inherent in the complete event ordering required to maintain strict coherence. Thus, they may yield greater concurrency in a multicomputer environment. Various mechanisms are considered further in Chapter 2, while Chapters 3 and 4 describe the binary semaphore based mechanism used in the implemented distributed object server.

## 1.2.2 Implementation Issues

The logical model defines the operations which applications can use to interact with the store. There are implementation issues common to most logical models which may significantly affect the store's performance, some of which are outlined below.

**system architecture**

Machines such as the AP1000 contain many independent processors, each with their own local memory. The question is: how to utilise this processing power to harness the distributed memory in a manner which efficiently supports the object store paradigm? Two possible system architectures are peer-to-peer and client-server. Chapter 3 compares these architectures and describes the implemented client-server solution.

**object structure**

Most existing object stores find it convenient to be able to distinguish between those parts of an object which are simple data and those which are references to other objects. This facilitates various optimisations such as *swizzling* (Chapter 2) and *prefetching* (Chapter 4). Several mechanisms exist for identifying references, the most common being to structure objects such that the data and reference parts are distinguished by a known boundary [9, 8]. This is the method used in the present implementation. Some systems use an *up-call* facility, in which application program methods are provided to identify the references in objects to which they are applied [24, 3].

**object migration**

The primary issue here is whether objects should migrate through the system to where they are needed, and have operations applied there, or whether methods should be invoked upon objects remotely. This issue is addressed more fully in Chapter 3. Assuming that objects do migrate, two forms of migration exist within a distributed object store; migration to and from stable storage, and migration between processors. Stable storage and issues related to it are beyond the scope of this project. Hence migration to and from stable storage is not considered further. The migration of objects between client and server processors is an issue which is investigated further throughout the thesis.

**accessing objects**

How do application programs access objects efficiently and safely? If objects exist within the application's address space they may be accessed directly or indirectly. Direct access is necessarily faster, but as is discussed in Chapter 3, it is not safe in a multicomputer based object store. Indirect access raises the question of how to organise an internal map to facilitate efficient look-up and access. This is discussed further in Chapters 2 and 3.

**replication**

The technique of allowing multiple copies of an object to exist in different locations may yield benefits from reduced object transport costs. The value of this technique is discussed further in Chapter 3 which also describes how replication is used within the constructed system.

## 1.3 Project Scope

The aim of this thesis is to explore issues and mechanisms relevant to an object store implementation on a multicomputer architecture. It details the design and implementation of a distributed object server — a distributed object store without the property of persistence. Most of the above mentioned issues are addressed in the description of the system's design and implementation. The issues of object migration and coherence maintenance are further addressed through an evaluation of techniques performed using example applications which utilise the implemented server. The distributed object server follows the client-server model, and different schemes are evaluated for migrating objects from clients to servers (referred to as space recovery), and from servers to clients (referred to as object migration). The client-to-server strategies evaluated are: memory dump, simple LRU, and a classified scheme. The server-to-client schemes investigated are: single object on demand, single object with prefetch, and migration by storage unit (segment). The coherence maintenance scheme evaluated uses binary semaphores on each object to guarantee mutual exclusion. The example data hungry applications used are primarily from the area of scientific simulation. These included two solutions to the $N$-body problem, and a plucked string simulation. The implementation's general performance characteristics are also reported. The objects of the system are both untyped and structured, consisting of explicit data and reference parts.

The distributed object server is implemented on the Fujitsu AP1000 [17, 18]. The AP1000 is a distributed memory MIMD machine. It possesses a communications network having processors placed at the intersection points of a two-dimensional torus grid. Processors are also connected by separate broadcast and communications networks. The machine does not support virtual memory or hardware address translation. The machine used for this work was configured with 128 processing elements, of which 32 had 500MB disk drives, and all possessed 16MB of memory.

The next chapter reviews previous work relevant to the issues mentioned above. Chapter 3 presents the logical model for, and design of, the system implemented. This is followed, in Chapter 4, by an outline of the system and a discussion of the data structures and algorithms used. Chapter 5 then describes the system instrumentation and example applications. The next two chapters detail the investigation of system performance, object migration schemes, and the coherence maintenance mechanism. Chapter 8 presents a discussion of the experiment results, and Chapter 9 details conclusions drawn from the project and discusses possible future work.

# Chapter 2

# Background

This chapter discusses some of the existing literature that is relevant in understanding the problem of implementing a distributed object store on the AP1000. It begins by outlining two important concepts in object stores, identity and referential integrity, and proceeds with a brief discussion of the design and implementation of PS-Algol[4] and the Mneme Persistent Object Store[27]. This is followed by a discussion of concurrency control and coherence measures, including a review of some existing methods and systems which may be useful references for an AP1000 implementation. The important concept of loose coherence is introduced and Munin[5], a system which exploits this, is reviewed.

## 2.1 Object Stores

Conceptually an object store is a space in which objects exist. An object being a group of related data items, possibly including references to other objects. Minimally, an object store should provide client applications with facilities to place new objects in the store, manipulate those already present, and to remove those no longer required. These facilities are however, dependent upon two more basic provisions — identity and referential integrity.

### 2.1.1 Object Identity

Implicit in the concept of an object store is the idea of object identity. We must be able to identify an object to access it, and this identity must be preserved for the duration of any access session that may occur during the life of the object. In many systems [26, 8, 27] this requirement is addressed by assigning to each object an identifier which is system-generated and unrelated to any name the client may associate with the object. This identifier is used by the system to manage links (discussed below) between objects, and functions chiefly as a pointer to the object within the store. Consequently, when an object is moved from the store to local memory, any references to other objects which it contains will not be directly valid as local memory pointers. These references must be translated to refer to the local memory address (if any) of their objects. Temporarily replacing object identifiers with local memory pointers is called *swizzling* [24, 42, 20]. Another less favoured option involves translating the required identifier with each access that occurs.

Kemper and Kossmann[20] have addressed the tradeoff of swizzling overhead against increased access efficiency. They have classified and investigated many of the existing swizzling techniques, concluding that no one is superior in all cases. They recommend that the choice be left to the application, or dynamically switched by the system to achieve best results. *Direct swizzling* is the replacement of an identifier with the main memory address of the referenced object, *indirect swizzling* replaces the reference with a pointer to a place holder object through which all accesses must occur. Kemper and Kossmann identify the expense of maintaining a *reverse reference list* if direct swizzling is used in an environment where objects may be removed from memory. This is avoided with indirect swizzling by simply marking the place holder object as invalid when the real object is removed.

Referential Integrity exists where the link between objects is persistent, immutable, anonymous, and not related to a naming scheme. A link with referential integrity guarantees that the object referenced will remain accessible for the lifetime of the link and, in strongly typed systems, that the type correctness of the link will remain correct[26]. When using name based references, known as associations, these properties are not guaranteed. Figure 2.1 shows how associations and links with referential integrity behave differently when a named object changes.

Another advantage of using links that possess referential integrity is that access is independent of any naming schemes (or lack thereof) which may cover the linking graph. The price is a loss of flexibility — decisions regarding which objects to access must be made prior to the link being created. With a naming scheme these decisions can be postponed, through the use of name changes, right up until the association is actually invoked.

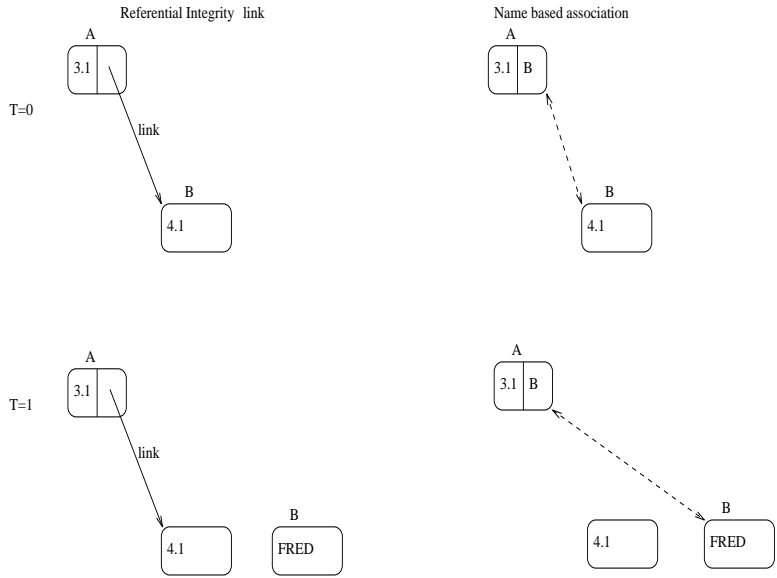

Figure 2.1: Referential Integrity compared to Naming Association as the referenced object changes between T=0 and T=1

## 2.1.2 Object Migration

Three object migration schemes were investigated during this project: single object upon demand, single object with prefetching, and migration by segment. The first scheme involves migrating individual objects as the need arises. The other two schemes attempt to predictively migrate objects in response to a single migration request. Migrating data by storage unit/segment is a well known data base optimisation technique which seeks to take advantage of a match between storage locality and temporal locality of reference. This technique has been explored in the context of object stores by Hosking and Moss[15]. They report speedups of up to 76 on benchmark applications when using segment migration rather than single object migration.

## 2.1.3 Recovering Store Space

The lifetime accessibility guarantee offered by referential integrity raises the question of when an object actually ceases to exist, that is: when is its memory deallocated? Two possibilities exist for determining that an object should be removed from the store. These are explicit deletion and garbage collection, both of which are outlined below.

The need to deallocate objects which are no longer required arises from the finite size of the store. The infinite capacity of the conceptual object store is simulated by the removal of unnecessary objects.

**Explicit Deletion**

A system using explicit deletion allows the client program to issue a specific command requesting the deletion of an object. The generally accepted view of this method is that objects cease to exist immediately the store is notified of the request. This however, raises the question of how to handle references to the deleted object which are still extent. Morrison *et al* [26] have proposed a form of explicit deletion which does not suffer this problem. For object stores using their method an object ceases to exist only when the last link to the object is removed. This is similar to the link command in UNIX[30]. Thus, whilst the object may no longer be reachable from a particular object whose link has been changed, links from other objects may still exist with the object reachable through them. This method is similar to reference counted garbage collection (see below).

**Garbage Collection**

If a system does not use explicit deletion then some form of garbage collection must be used to reclaim storage allocated to objects that are no longer required. Garbage collection may occur either automatically or at the behest of the client. It is performed by removing all objects which are no longer reachable from predetermined root objects[8]. Care needs to be taken regarding any object references that active clients hold at the time of garbage collection. It is possible for the client to possess a reference to an object which is unreachable from any root object at the time of garbage collection but which the client intends to make reachable using the reference it possesses. If this is allowed to occur, the garbage collection will deallocate the object and the client's reference will no longer be valid. A solution is to include the environment of each process in the set of root objects from which reachability is determined. Figure 2.2 demonstrates how an object becomes a candidate for deallocation under garbage collection.

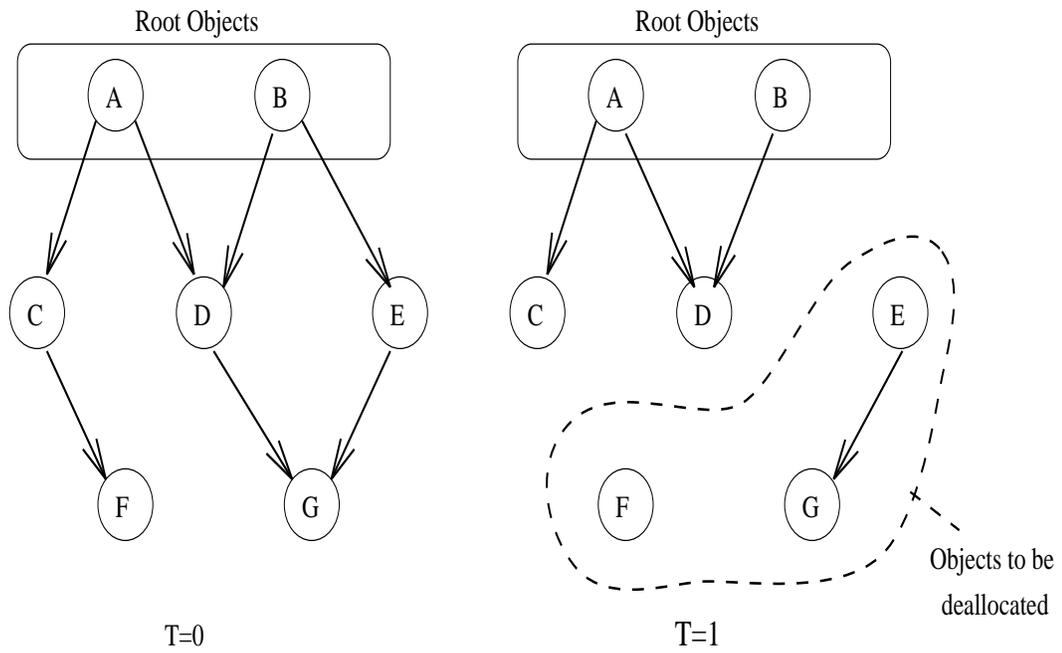

Figure 2.2: **Garbage Collection** : if the links BE,CF, and DG are broken between T=0 and T=1 then objects E,F, and G become candidates for deletion

### 2.1.4 Extant Systems

**PS-Algol**

PS-Algol is a version of S-Algol to which a persistent object store has been added by teams from the University of St Andrews and the University of Edinburgh[26, 4, 35, 9]. From this project Atkinson *et al* raise many design issues and present some of their solutions.

By their nature, object stores are generally separate entities from the client programs they service, that is: they conform to a client-server architecture. The object store consists primarily of a server (a separate process [24, 32] or run-time support [4, 27]) and possibly an interface library to facilitate client requests of the server.

Atkinson *et al* raise several design issues that deserve consideration —

- Should all references go to the central store? Alternatively, should some objects be in local storage and a lookup table be used instead?

- At what time should this lookup table be built? Are entries to be made as references arrive locally inside other objects, or when the reference is itself first used?

- How can the map be stored with low overhead but sufficient space and performance?

- What about garbage collection?

- Virtual memory?

- What address structure is extensible, space efficient, and can address a sufficiently large space?

- Placement strategies?

- Transaction implementation?

When contemplating an object store on the AP1000 some of the above concerns are extraneous — virtual memory does not exist at the cell level for instance. The findings of Atkinson *et al* are, however, of considerable relevance.

They began with a linear table (PIDLAM) mapping Persistent Identifiers (PIDs) to local addresses (LAs). Although hashing was used to accelerate access, performance was less than desired. In another implementation, the position in the PIDLAM was referred to as the Local Object Number (LON). This was used for all local references, costing one indirection for all accesses but giving good performance on local reference to PID translation (simply indexing into the table). This method required that every active object have an entry in the PIDLAM, including non-persistent (that is, non-store resident) ones[1], and consequently the table became very large. A third incarnation has abandoned the LON. It has adaptive structures to avoid high fixed overheads or small limits, and a PIDLAM consisting of two hashing structures — one from PIDs to LAs, one from LAs to PIDs. To minimise table size only those PIDs that have been imported and dereferenced are included in the PIDLAM. Garbage collection occurs automatically when transactions are committed and data exported to stable storage (that is: when it is returned to the central store).

### The Mneme Persistent Object Store

The Mneme project[27] includes a more generic implementation of an object store than PS-Algol. It seeks to support cooperative, information-intensive tasks such as Computer Aided Design (CAD), by providing a distributed persistent object store.

In order to avoid the communication and operating system overheads incurred when running a client-server architecture, most Mneme functionality has been included as either a linked library or run-time support depending on the particular application. Whereas the primary abstraction supported by the client interface is the object, the primary abstraction for the client-server interface is the *physical segment*. A physical segment is a vector of bytes generally containing many objects and thus providing clustering for storage and retrieval. Objects within a physical segment are accessed using a server specified label which is not guaranteed persistent across sessions (due to reallocation after garbage collection etc). The relationship between objects (that is: the referential integrity of the links) however, is guaranteed persistent.

Mneme allows multiple *files* (nominally independent groupings of objects) which contain several *pools*. A *pool* is a group of objects within a unit that is subject to a management policy independent of other pools, and whose objects are distributed over one or more whole physical segments. Mneme objects can be accessed either via the defined client interface functions (using *handles* which are similar to *capabilities* used in the acacia file system on the AP1000[7]), or directly through pointer manipulation. If using pointer manipulation, the client programs must take care to inform Mneme about modifications made to objects, and to not dereference unswizzled references.

---

[1]It must be remembered that Atkinson *et al* were working on an object store embedded within an existing language and to provide two access methods for objects would have been infeasible.

Object identifiers (references) in Mneme are 32 bits long and valid only within their own file. Objects located in other files may be referenced using *forwarder objects*. These are a local proxy for the foreign object and invoke predefined forwarding routines when referenced.

Whilst the client interface presents an object based abstraction, the server level abstraction, in which objects are moved only as part of an entire physical segment is very similar to paged virtual memory. In a multicomputer environment this may be expensive, since it risks the movement of unnecessary data and raises the possibility of false sharing[2].

## 2.2 Distributed Systems

The degree of concurrency in the AP1000 is far greater than that managed, or even considered, by any of the systems mentioned in the previous section. Indeed, from the concurrency viewpoint an object store for the AP1000 will have much in common with distributed virtual shared memory systems, and will require concurrency control and coherence facilities well beyond those inherent in most existing object managers. The problem to be solved here is how to handle multiple updates from different processors in such a way as to minimise the coherency operations and obtain maximum efficiency from the processors.

Two existing techniques for handling concurrency in object stores, two-phase locking and optimistic scheduling, are first considered. The concept of loose coherence is then introduced, and an existing system that emulates object based shared memory on distributed memory machines is reviewed.

---

[2]*False Sharing* occurs when processors in a shared-memory parallel system make references to different data objects within the same coherence block (cache line or page), thereby inducing unnecessary coherence operations.[6]

### 2.2.1 Two-Phase Locking

The concurrency control method employed by most object managers and database management systems is simple two-phase locking. This method is used in the Cricket[32] and Cool[24] systems, and has been proposed for Mneme[27]. Two-phase locking is explained by the following simple rules relating *transactions*, locks, and objects. A transaction is a set of object updates which together move the store from one consistent state to another.

- In order to update an object a process needs to do the following in order;

  1. Obtain the object's write-lock
  2. Update the object
  3. Commit the update to the store
  4. Release the write-lock

- A process cannot obtain an object's write-lock whilst another process possesses it — usually blocking until it becomes available.

- No lock obtained during the transaction is released until all locks required have been obtained.

- A transaction will fail to commit if deadlock occurs, and the system detects it.

- If the transaction fails to commit the object manager will roll back the update and release the write-lock.

The third rule guarantees serialisability[3] by implying that all locks precede all unlocks[16]. The last rule requiring the update to be all-or-nothing. Updates are hence serialisable and correct.

Two-phase locking can be made to work in a distributed system, but this requires locking all copies of the data to be updated. Such an approach is only feasible where updates are relatively infrequent compared to the response time, and the update processing does not block other processes for unacceptable periods of time. Where a 5 second update processing time may be acceptable in an office performing 2-3 updates an hour, even a 0.2 second update processing time would be unacceptable on the AP1000 if updates were desired at a rate of 20-30 per second.

---

[3]If a sequence of operations of a set of concurrently executing transactions is such that their effect on the store is the same as if they had been executed sequentially in some order, then the transactions are said to be *serialisable*. This is the criteria by which correctness of concurrent execution is judged [16].

The AP1000 is a machine built for high performance, this is gained through a high degree of concurrency. Having processes blocking whilst waiting to perform updates reduces concurrency and hence performance on machines like the AP1000. Another drawback of two-phase locking is that, where it is used in an environment containing two or more lockable objects, it is possible for deadlock to occur[16].

### 2.2.2 Optimistic Scheduling

Optimistic scheduling [16] is an approach that avoids the system overheads inherent in two-phase locking by not performing any locking. The system works by allowing all processes access to data objects, with any updates being written to local "shadow" objects. An associated record of both the read set (data read by the transaction) and the write set (data updated by the transaction) is also kept. Before the transaction is committed to the store it is checked for validity with respect to the transactions which preceded it. In order to be valid, a transaction T must satisfy at least one of the following criteria with respect to $T_i$, where $T_i$ denotes any transaction preceding T:

- $T_i$ committed before T started.

- $T_i$'s write set does not overlap with the read set of T and $T_i$ completes committing prior to T's commencement of committing.

- $T_i$'s write set does not overlap with the read set or the write set of T.

If the previous transactions were serialisable and T satisfies any one of these criteria then the set of transactions containing T and the previous transactions is serialisable. If T fails to satisfy at least one of these criteria then the transaction is aborted and the store is rolled back (restored to its previous state).

In this manner optimistic scheduling basically abandons locking and wagers that the transactions will be serialisable. If the scheme wins the wager it has avoided the overheads of locking, if it loses it has to roll back the non-serialisable transaction and either attempt it again or abort it. Thus, although the scheme avoids the possibility of deadlock, it incurs overheads in both time and space because it must maintain and then validate shadow copies and associated records of the read and write sets.

Perhaps the biggest disadvantage of this scheme in the context of an AP1000 object store would be the penalty for a non-serialisable transaction. If a cell's update fails to be serialisable, it could be expensive to force the cell to repeat the transaction.

### 2.2.3 Coherence - Strict & Loose

Two-phase locking and optimistic scheduling both enforce strict coherence on the store, that is, the value returned by any read operation on an object is the most recent value written to that object by any process. A parallel program is generally only a partial ordering of the events within the program, the partial ordering being specified by explicit synchronisation and implicit knowledge of other threads[5]. The synchronisation operations necessary to achieve strict coherence impose a more complete ordering at run-time. In terms of performance, however, synchronisation operations can be expensive on multicomputers and are thus prime candidates for minimisation. What is needed is a method of maintaining coherence without imposing the more complete, but costly, ordering required by strict coherence. This is not possible with our strict definition of coherence, but if we relax our definition somewhat then a weakening of the ordering can occur yielding fewer extra synchronisation operations. Figure 2.3 demonstrates, by way of a simple example, how these two types of coherence differ.

> **loose coherence** — the value of a read operation on an object is the value written by an update operation to the same object that *could* have immediately preceded the read operation in some legal schedule of the threads being executed.

With loose coherence it is possible to delay notifying other threads of object updates until they could otherwise have determined that an update occurred, such as through communication or synchronisation. When two processes communicate or synchronise there is an implicit acknowledgement by each that they are at a certain point in their execution and the other process is fully justified in making assumptions based upon this. For example, if two processes, A and B, are executing and B updates an object, then if no subsequent communication or synchronisation has occurred, A cannot infer that B has performed the update because B may not have progressed to that point in execution. Once A and B communicate or synchronise, A can infer that B has arrived at the communication point in its execution and has hence also done the update. Communication or synchronisation between two processes can be direct, as just outlined, or indirect, such as through one or more tertiary processes. For example, if process B communicates with process C, C then knows where B is up to. If C then synchronises with process A, A will implicitly acknowledge A C's stage of execution and can thereby infer that B has at least progressed to the stage of communicating with C.

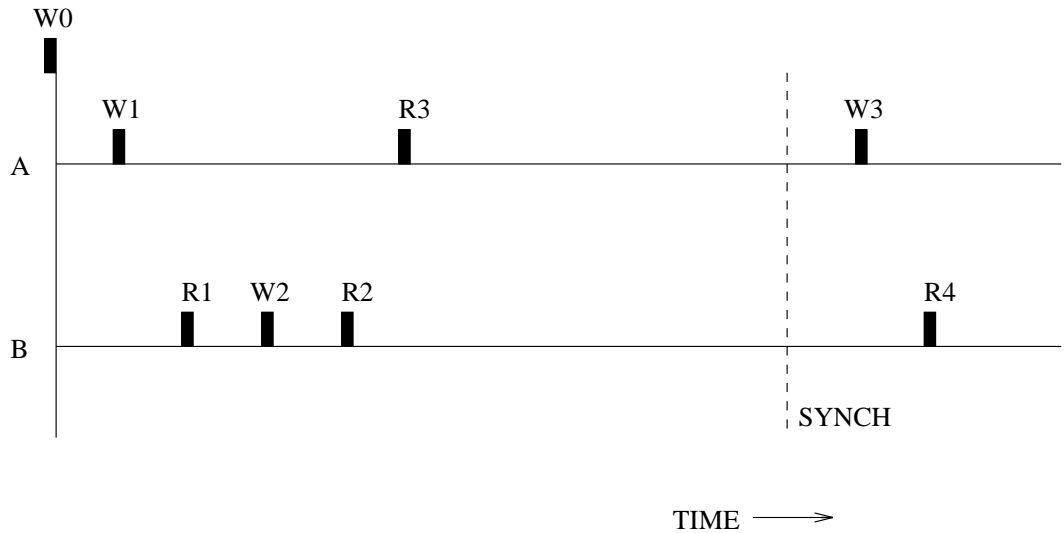

Figure 2.3: — **Strict & Loose Coherence** : A and B are separate processes, with W0 through W3 being writes to an object, and R0 through R4 being reads of the same object. Assuming W0 is an initialisation write at program execution then under strict coherence R1 would read the value written at W1 (it being the most recent value written), R2 would read that written at W2, as would R3, and R4 would read that written at W3, the synchronisation having no effect. Using W0 as the same initialisation write, under loose coherence R1 could read either the value written at W0 or W1 (W1 is in process A and without communication or synchronisation process B cannot infer that W1 has occurred), R2 must read the value written at W2 since it logically precedes the read in the same process, R3 may read either the value from W1 or W2, and R4 may read any of the values written at W1, W2, or W3 (W3 occurring after the last synchronisation cannot be inferred to have occurred, the occurrence of W1 can be inferred from the synchronisation, and W2 is known to have preceded R4, but there is no way of knowing which of W1 and W2 occurred most recently).

One disadvantage of loose coherency is that it may be less intuitive for programmers to reason about. Programs run under loose coherence are semantically the same as those run under strict coherence[4]. This is because under strict coherence a programmer cannot make assumptions about the execution progress of various threads between synchronisations. However, an intuitive (and often correct) feel for the progress is sometimes used.

Loose coherence guarantees that the order of visible updates is preserved for all processes, at the cost of more expensive update commitments. This does however, avoid the rollback expense sometimes incurred by optimistic scheduling. Loose coherence has the potential to significantly reduce the coherence related synchronisation overhead, and as such could provide a notable performance boost for multicomputers running concurrent update applications.

### 2.2.4 Extant Systems

**Munin**

> Programmers want shared memory. [31]

This is what motivated Bennet, Carter, and Zwaenepoel [5] in their construction of Munin, a distributed shared memory system. Munin seeks to bypass the architectural difficulties of hardware supported shared memory systems by allowing shared memory programs to execute efficiently on distributed memory machines. Performance on the distributed memory machine is maintained through the use of type-specific memory coherence and by exploiting loose memory coherence.

---

[4]Programs using access to shared objects as an implicit means of communication or synchronisation would appear to invalidate this statement. Such programs can, however, be rewritten to use explicit communication and synchronisation, thus maintaining the statements validity.

Bennet *et al* conducted an investigation of the object access behaviour of parallel programs at the programming (hence architecturally independent) language level, and found the following to generally hold true:

- Few objects are both consistently accessed and have reads and writes being of the same magnitude.

- Programs have several different phases of execution, with each exhibiting an access signature, for example: Initialisation, Computation, Result Assembly.

- The majority of accesses are reads, except during initialisation.

- The mean time between synchronisations is far longer than the mean time between accesses to shared data objects.

The first 3 of these generalities lend support to the possibility of different coherence measures for different classes of objects, whilst the last indicates that the exploitation of loose coherence could reap a substantial performance saving.

An additional contribution of the parallel program analysis mentioned above is the following categorisation of objects by access type:

- Write-once objects

- Private Objects

- Write-many objects

- Result Objects

- Synchronisation Objects

- Migratory Objects

- Producer-Consumer Objects

- Read-mostly Objects

- General Read-write Objects (the catch-all case)

Munin currently runs on an Ethernet network of SUN workstations with each machine running a Munin server interacting with the client application and the underlying operating system. When a client application accesses an object for which there is no local copy a memory fault occurs and the server invokes the appropriate handler for the object based upon its access class. It is through using these class specific handlers in the propagation of updates that Munin exploits loose coherence. Each thread maintains a list of updates yet to be propagated in a *delayed update queue*. When an object is updated, the server for the updating process enqueues a record of the update on the delayed update queue. When the thread synchronises, the queue is emptied by broadcasting the update records to the servers possessing copies of the objects updated. Remote servers acknowledge receipt with a message indicating that their local copy was invalidated or that they no longer hold a copy of the specified object, thus allowing the local server to update its distribution list.

As noted, each access class has its own specific handler which implements coherence measures appropriate to the class. These measures are tailored to suit the update notification needs of the class in light of loose coherence, and are derivatives of the more general case outlined above. Consider the following examples of Munin's coherence tailoring.

**Write-once** objects are written during the initialisation phase and thereafter are only read. Munin supports this class by simple replication. After the initial write no updates occur and so no specific coherence policy is required.

**Write-many** objects are frequently modified by multiple processes. This is supported by object replication and the delayed update mechanism mentioned above.

**Result** objects are essentially a sub-class of Write-many objects, and like that class utilise the delayed update procedure mentioned above. However, Munin further optimises this by seeking to take advantage of the fact that as result objects are not read until the data is collated, updates to different parts of the object will not conflict.

**Read-mostly** objects are replicated by Munin which performs in-place updates via broadcast. The use of broadcast is justified by the infrequency of updates.

Munin presently relies upon programmer provided semantic information to classify objects, but the system's developers hope to acquire some of this information automatically at compile time. They have also considered the possibility of run-time determination of object class.

The method of delayed updates allows Munin to combine updates to the same object, and also to piggyback update propagation onto synchronisation based communication. Hence, the amount of data sent, and the number of data items sent are both reduced. This represents a potentially significant performance boost over an equivalent system based upon more synchronous coherence methods such as two-phase locking mentioned earlier. Performance is important for applications on the AP1000, so the techniques employed in Munin may well warrant further consideration with regard to an AP1000 object store.

## 2.3   Overview

The breadth of literature relevant when considering object stores on a multicomputer is considerable. It encompasses such fields as persistent object stores, database management, network communication protocols, and shared virtual memory.

Irrespective of whether it is persistent and on a uniprocessor, or non-persistent and distributed across a multicomputer, an object store is still a space in which objects are kept. As such, it is built upon the basic concepts of object identity and referential integrity. With no object store being infinitely extensible a method of deallocating useless objects is necessary, this can be either through explicit deletion or garbage collection. Much of the house-keeping information within an object store is in the form of look up tables. Atkinson *et al* [26] provide valuable insights into possible structures to fill this role. Their final implementation being a PIDLAM utilising hashing techniques on two adaptive structures, each the inverse of the other. Mneme [27] presents innovation in the form of object groupings with files containing pools having individual management policies, and forwarder objects for relations between objects in different files.

The most common concurrency control methods, two-phase locking and optimistic scheduling, both guarantee serialisable transactions. Neither is considered ideal for high performance applications on a multicomputer. Loose coherence has the potential to provide better performance on message passing parallel machines, especially when combined with lazy propagation protocols such as Timestamped Anti-Entropy developed by Golding and Long[14]. The idea of type-specific update notifications has been introduced in the Munin[5] distributed shared memory system, and has the potential for achieving even better performance through the tailoring of update propagation techniques.

# Chapter 3

# A Multicomputer Object Server

In this and the next chapter the design and implementation of an experimental distributed object server for the AP1000 is described. This system was constructed both as an aid to understanding how an object store might work on a multicomputer, and to facilitate the comparison of different techniques and data structures when utilised in a multicomputer object store. The aims of the system are discussed further in the following section, and some definitions are laid down providing a framework for a discussion of how best to utilise the MPP style hardware. The software architecture of the system is then given. The next chapter continues the implementation discussion, detailing data structures, the major store operations and the different techniques compared.

## 3.1   Implementation Aims

The implemented distributed object server is an experimental system, and has two classes of aims. It must obviously seek to satisfy the aims of a distributed object store, but must also do so in a manner which is flexible enough to facilitate the experimentation necessary for comparisons between different techniques. There would be little point in constructing an experimental system which achieves the object store goals very well in some particular manner, but is so constructed as to preclude any modification which would enable comparative techniques. The aims of object stores were mentioned in Chapter 1, but to state them explicitly with regard to the implemented object server, the system seeks to:

- provide an object-oriented memory paradigm

- enable processes to work with data sets larger than their local memories — virtual memory

- manage concurrent access to data objects, allowing sensible semantics — sharing

As with all things in computing, speed and transparency are also highly desirable. Hence, the implementation goal is to provide, in a flexible manner, fast and transparent access to an object-oriented, shared virtual memory.

## 3.2 Definitions

This section defines some concepts basic to the object server implementation. Several of the terms have been discussed in previous chapters, often in a general sense and sometimes with regard to specific systems. They are redefined here to provide a clear and consistent framework for the description and discussion of the experimental system.

### 3.2.1 Objects & Object Identifiers

An *object* is conceptually represented by the triple:

$$< \underline{object\_identifer}, data\_part, reference\_part >$$

If the entire object store is considered as a single table in a relational database, the *object_identifer* is the primary key. Hence, for each object/row this field is unique, non-null, immutable, and is the means by which objects/rows are selected for retrieval or update. Unlike keys in most relational databases[1], *object_identifiers* are system allocated and should not be constructed by the user. Pointers in C or Modula-2 which are used to identify data structures are analogous to the *object_identifiers* used here.

The *data_part* of an object consists of a single sequence of bytes in which may be stored information. The information stored here does not include references to other objects, but the value of the object itself. The length of the byte sequence making up the *data_part* is a property of the individual object and is determined at the time of object creation. The storage units called *segments* (discussed below) which are used by the system, impose an upper bound on this length.

---

[1]Lorie and Plouffe [25] described extensions to IBM's System R relational database product which allows it to better manage complex objects and their relationships. One of these extensions involves the system allocation of unique primary keys to identify objects.

The *reference_part* of an object is a sequence of references to other objects. As with the *data_part*, the length of this sequence is determined at object creation. A reference to another object takes the form of an *object_identifier*, hence the *reference_part* is a sequence of other objects' *object_identifiers*. Just as it is necessary to distinguish between valid and NULL pointers in C, it is necessary for programs using the object server to be able to determine which references are valid and which are not. For this reason, the constant reference known as NULL_REF exists and can be used for comparison in the same manner that NULL is used for pointers. The server system internally also uses a low level constant called NULL_OID.

### 3.2.2 Segments

Whilst the paradigm presented to the user is that of an object-oriented memory, the implemented system (like many others) stores and manipulates objects within its own fixed size unit of storage. Here the unit is called a *segment* and it is the size of the segment which places upper bounds on the sizes of individual objects. No single object can be larger than a segment, equivalently the sum of the *data part* and *reference part* of any one object cannot exceed the size of a segment. A segment may however contain many objects, with the system using a *directory*, one of which is associated with each segment, to locate the objects.

It is hoped that some locality of reference can be achieved by transferring not single objects but entire segments at a time. Thus, a segment needs to be of sufficient size not only to store any object which we may reasonably expect to create but, if any locality of reference is to be gained, to also store its neighbours. Each segment is identified by a *segment identifier* which is unique within the *group* (defined below) to which the segment belongs.

### 3.2.3 Group

From the user perspective, objects are members of groups. A group is a set of object management strategies and is created by selecting from the strategies supported by the system in each of several categories. This is done at the time of the group's creation and, under the current specification, fixed thereafter. A group is known by its *group identifier* which is generated at the time of creation. The group to which an object belongs determines which set of object management strategies are applied to the object. The following are examples of object management strategies:

- moving objects individually or moving entire segments

- whether to prefetch objects, and if so to what degree[2]

- across how many servers to spread the objects of the group[3].

Differing object management strategies may well affect the efficiency of an application using the store, but do not affect the semantics of any operation upon the store.

When programming, it is often the case that there are many data objects which are similar in size and structure and which are accessed not only in similar patterns, but are also temporally clustered. The nodes of a linked-list, or of a binary-tree, are such structures. The group is seen as a way of tailoring the system's management strategies to provide better performance when working with such objects. Since it is likely that such objects are to be accessed close together temporally, it is logical that they be close spatially as well. This produces benefits from locality of reference, via the mechanism of transferring entire segments at once. For this reason, all objects belonging to a group are stored in segments which are also said to belong to the group and which contain only objects of that group. Hence, if the nodes of a binary-tree all belong to the one group, they will also all be stored in segments belonging exclusively to that group. If the object management strategies for the group are appropriate, they will combine with the spatial locality of reference to yield better performance than would otherwise be the case.

## 3.3  MPP hardware — how best to use it?

As noted in Chapter 1, machines such as the IBM SP2, Cray T3D, CM5, and AP1000 have certain architectural features in common. These features distinguish MIMD MPP machines as a class apart, and include:

- a fast, reliable communication network between processors

- a large total memory distributed across these processors

---

[2]This is discussed more fully in §4.7.
[3]See §3.4.4 for further discussions of the implications of this strategy.

- many processors, each capable of independent computation

Implementing an object store on such a machine raises the question of how best to take advantage of these characteristics. If we are unafraid to make generalisations and temporarily exclude hybrids, there are essentially eight ways to utilise such systems. These eight possibilities stem from binary choices made on the utilisation of each of the above three characteristics in response to the following questions:

1. does the network communicate objects or operations upon objects?

2. can more than one copy of an object reside in the total memory?

3. do the processors all perform the same object management tasks?

The fast communications network is what most distinguishes MPP machines from clusters of workstations, and it is the utilisation of this network which is considered first.

### 3.3.1 Utilising of the Network

In the object-oriented domain, two paradigms exist for the network. It can either be a transport network or a communications network. As a transport network, it is used to shift objects (or their replica's) around the system to where they are needed. If used as a communications network, messages are sent to and from objects which do not move around the machine. These messages may cause the object to change state, and play a similar role to method invocations on C++ objects. A real world analogy to these paradigms is to consider the road system as a transport network and your phone/fax/email system as a communications network.

Since we are looking for an object store which will be useful in high performance applications, we need to ask which utilisation is the most efficient. In other words, which will incur less cost in the long run? If we look purely at the number of messages required by each paradigm we see that given $r$ read accesses and $w$ write accesses to an object, the communications network would require $w + 2r$ messages[4] whilst the transport network could require up to $2(w + r)$ messages[5] assuming serialised access and no replication. Several factors complicate further analysis along this line, notably different message sizes required in each paradigm, the processing requirements of operations on the objects, and the amount of memory each process has available to store objects under the transport network paradigm.

---

[4]This assumes that all access are external to the object's process.

[5]One message to request the object be sent, and another to actually send it.

Transporting large objects is expensive, and would become a significant overhead for the transport network if many were accessed by processors in such a way as to require frequent movement.

Operations on objects are performed by the processor "owning" the object under the communication network paradigm. Since no existing MPP machine has enough processors to allocate a single processor per object for a reasonably sized object store, many objects would have to be "owned" and therefore serviced by a single processor. If several of these objects are to be accessed at once, efficiency is lost as the accesses must be serialised (or executed concurrently at the expense of increased context switching) by the owning processor. The transport network does not suffer this problem because objects migrate to the processor wishing to access them and which is subsequently responsible for performing the operation.

The memory available to a processor for the storage of objects which are transported to it may also impact upon the cost of using this paradigm. If processors access so many objects that they gather a collection larger than their locally available physical memory, the expense of coping with the problem may be significant. Actions such as migrating objects to secondary storage or other processors, will solve the problem at the expense of considerable overhead. This is especially so if these objects are subsequently accessed again requiring their placement in memory and the removal of other objects.

### 3.3.2 Object Replication

Turning now to the question relating to the machine's memory, the issue is whether or not multiple copies of a particular object will coexist in the memory of the machine. As with all computers, the memory on the class of machines being considered is never enough, and any decision to consume significant portions of the store with replica objects will require justification. No advantages are apparent if the system utilises the communications network paradigm outlined above. If the transport network is adopted then it is possible that object replication will provide benefits. The idea is simply that if a process other than the holding process wishes to access an object it will request it, as is done in the previous section. However, the holder need only transport the object to the requestor if it has been updated by some process since the requestor last held it, or the requestor has not previously
held the object. This assumes that a process does not delete its local copy
when transporting an object to a requestor, and that there is some mech-
anism for determining if an object has been updated since last held by a
particular process.

### 3.3.3  Peer-to-Peer vs Client-Server

It is also necessary to consider the allocation of the object management tasks
over the processors. Two paradigms are obvious, these being the peer-to-
peer and client-server models.

Examples of the peer-to-peer model include Munin[5] and Global Array
Toolkit [28], both of which are distributed shared memory systems relying
on tight coupling of processes. The overheads, both memory and processing,
inherent with coalescing the management tasks and user tasks into the one
process remain unreported for these systems.

As noted in Chapter 1, conventional object stores such as Mneme[27] are
based on the client-server model. These have specific server processes man-
aging the store, and routines in the client processes managing local memories
and communicating with the servers as needed. This model does not require
the tight coupling of the peer-to-peer model, and if clients and servers are
run on separate processors, entails no overhead through competition for
resources.

ANU Linda [10] demonstrates that these two models are not mutually
exclusive on the AP1000. In this system, each AP1000 cell runs two tasks,
a tuple server and a user task. Data objects (*tuples*) are distributed across
all cells using a hashing function. When a match for a tuple is sought,
communication must occur with all cells possibly containing matches (as
determined by the hashing function). This is similar to the case of non-
replicated object migration discussed earlier. Experience has shown the
overheads inherent in this model to be significant, requiring careful crafting
of code for adequate performance.

## 3.4  System Overview

The system implemented follows the client-server model with each process
(server or client) executing alone on an AP1000 cell. Object migration
is the mechanism for providing coherence across concurrent accesses, with
replication and time-stamping used to improve performance.

The architecture of the system is shown in Figure 3.1. The stable object store shown is not part of the experimental system constructed, rather a conceptual expansion. The dashed box bounds the system implemented and investigated.

The client-server model decomposes neatly into two distinct software components. From a software engineering perspective these components are unwieldy. The client-side component is required to interface with the client program, manage the local memory and objects contained within it, and also to communicate with the servers. On the server-side it is necessary to communicate with the clients and other servers, and manage the server memory's contents. These components are broken down into a layered architecture as shown in Figure 3.2.

### 3.4.1 Application Programming Interface (API)

This layer provides the front end functions, data structures, and operators used by client programs. It aims to present the shared distributed heap in a manner such that the management tasks performed by the lower layers are transparent. Objects can be allocated and manipulated on the shared heap in a fashion similar to that used for a transient heap.

The distributed object server makes available two data types that application programs can directly create, manipulate, and destroy. These are the object reference and group identifier types. Instances of either type have values which identify an object or an object group respectively. Certain API operations return instances of these types which can be assigned to or compared with variables of the appropriate type. Operations available in the API include:

#### Client Identity

It is convenient for each client process to possess a system allocated identity. This facilitates the partitioning of work amongst different client instances. This operation returns the identity of the client, as an integer in the range from zero to the number of clients less one.

#### Create Group

This operation takes a movement context (SINGLE or SEGMENT), a width, and a prefetch depth, inserts appropriate defaults if width and/or prefetch depth are omitted, and returns a newly allocated group identifier. This identifier must be assigned to a variable of appropriate type if the group created is to be used when creating objects.

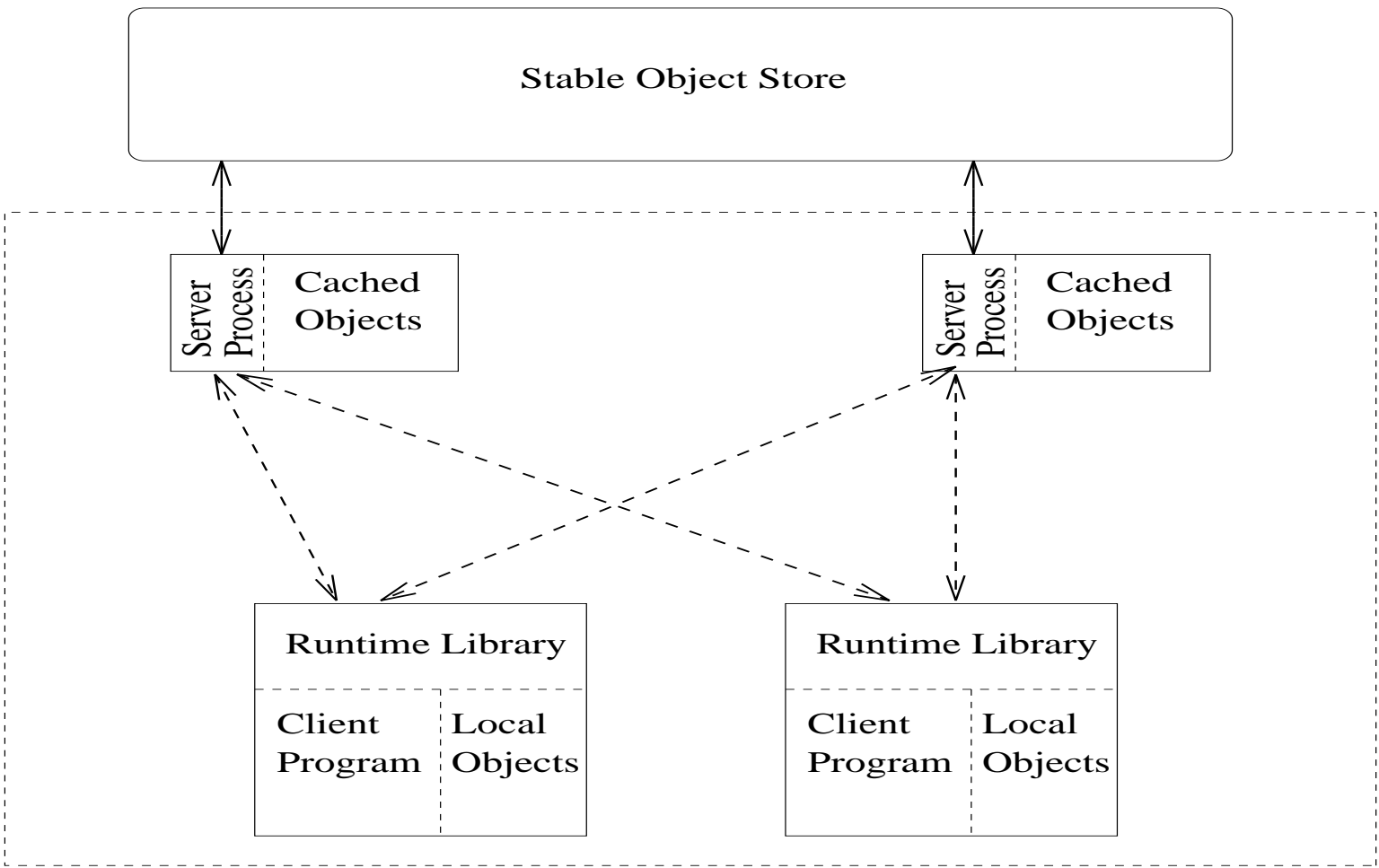

Figure 3.1: System Overview showing two clients and two servers

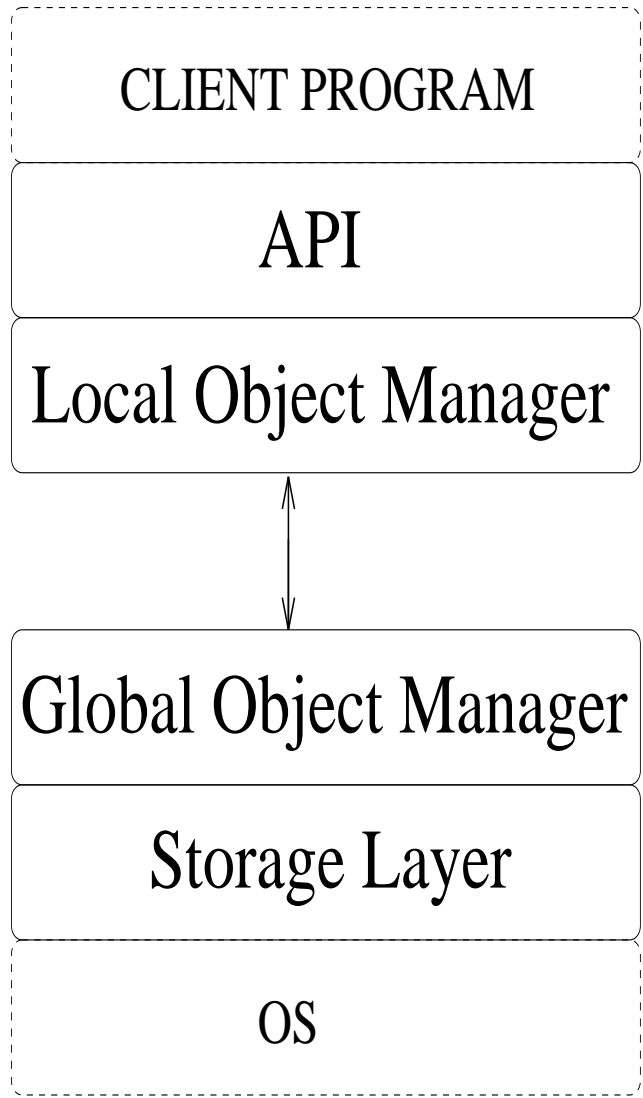

Figure 3.2: The Layered Architecture

### Create Object

This operation requires that three parameters be supplied. These are a valid group identifier, a size (in bytes) for the data portion of the new object, and the number of object references to be contained within the object. A reference to the newly created object is returned. This reference must be assigned to a variable of the requisite type if the object is ever to be accessed.

### Dump

This operation clears the local replica cache. This complete cache flush has the side effect of ensuring all objects so far updated by the client process are written back to the servers. With the updates thus available to other clients, this operation can be used in conjunction with the synchronisation operation (below) to obtain a crude form of coherence maintenance for concurrently accessed objects.

### Synchronise

This is a blocking operation which will not return until all other clients are similarly blocked, and all store operations (such as dump detailed above) initiated by any client have completed.

The majority of the other available operations are methods of the object reference data type. In addition to the following methods, assignment and equality comparison are also supported.

### read

This operation is used to read the data contained in the object identified by the object reference upon which it is invoked. It copies a specified amount of data from a given offset in the object's data part, to a given transient memory address. This involves accessing the object through the LOM as described below.

**write**

This operation copies a specified amount of data from a given transient memory address to the data part of the referenced object, beginning at a given offset.

**indexing**

Object references contained within other objects are accessed by indexing into the containing object. This is achieved by applying an index operator to an object reference (the containing object), with the index (a positive integer) supplied as an argument. The operator yields the object reference contained within the original object and associated with the index. This object reference can be used on either side of the assignment, equality, and inequality operators.

**Wait & Signal**

Associated with each object is a binary semaphore which can be utilised to guarantee exclusive access by a client. The Wait operation blocks until the associated semaphore is non-zero, then decrements the semaphore and returns. The Signal operation increments the semaphore and returns immediately. Any updates made by a client are propagated to the server when the Signal operation is performed, and any changes made by other clients are seen by a client returning from a blocking Wait. It is an error to Signal an object's semaphore in the absence of a preceding and corresponding Wait.

**object naming**

For different client processes to access the same object, they must each possess a reference to the object. The system facilitates this by maintaining a system wide name/reference mapping. Any client process may associate a name with an object reference, this overwrites any previous associations with that name. Clients may retrieve the references associated with names, and assign these to references they possess. If an attempt is made to retrieve a reference for a name for which no association has been made, NULL_REF is returned. A name is a null-terminated string.

### 3.4.2   Local Object Manager (LOM)

This layer implements the major operations required on the client-side and invokes management tasks, such as recovering space and fetching objects, automatically. It is responsible for managing the portion of local memory allocated to caching replicas of objects and communicating with the servers as required to perform this management and the operations requested via the API.

The major operations provided by the LOM to the API are:

**Group Creation**

To create a group the LOM accepts the parameters describing the group's object management strategies and sends them as part of a request to the *main server* (an arbitrarily chosen server responsible for performing non-distributed management tasks). The main server allocates a group identifier and a block of servers (defined by an allocated *base server* and the width supplied by the API) to service the new group. It then returns the information to the LOM which adds the new group to its *group table* and returns the identifier to the API.

**Object Creation**

To create an object the LOM accepts the group and size parameters from the API, ensures enough free space is available to hold a replica of the object and sends a request to the base server of the identified group. One of the servers responsible for the group then responds with a newly allocated object identifier. If the group's movement context is by segment and the client does not possess a replica for the segment in which the object was created, one is sent. If a replica segment is not received, the object is created in local memory or in a replica segment already possessed. The object is inserted into the resident object table (ROT) and the allocated identifier is returned to the API.

**Swizzling**

Before the API can execute any methods on an instance of the object reference data type, the object reference must be *swizzled*. Swizzling a reference temporarily changes its contents from the object identifier it holds whilst in the servers, to a pointer to the resident object table (ROT) entry for the object identified. This is achieved using the find object operation detailed below.

### find object

This operation searches the ROT for a given object identifier and returns a pointer to the entry once found. If an entry for the identifier is not found, then one is created, a replica fetched from the servers, and a pointer to the new entry returned. The automatic fetching of a replica in this case occurs because the absence of an entry indicates that no replica is present, whilst the call to find object is an indication that the object is about to be accessed.

ROT entries are reference counted, and will not be deleted whilst there are object references still holding pointers to them. Hence, only one find object operation need be executed per object reference.

### object access

The API gains access to the data objects through methods invoked on their corresponding ROT entries. These methods not only perform the required operations of read, write, Wait and Signal, but also invoke the fetch object operation if the object being referred to does not have a replica in local memory at present.

### synchronisation

This is achieved by broadcasting a synchronisation request to all servers and blocking until all servers have acknowledged the request.

### name object

This operation registers a name/object identifier association with the main server. The API supplies the name and an object reference. The object identifier is either extracted from this reference, or from the associated ROT entry.

### object named

This operation contacts the main server requesting the OID associated with the name supplied by the API. An object reference is constructed from the OID returned by the LOM, this is then returned to the API.

Management tasks which may be automatically invoked whilst performing the above operations are:

### space recovery

The LOM manages a fixed amount of memory. Within this memory it places the ROT, the group table, replicas of objects or segments, and other miscellaneous management information. When a request for memory for one of these items is made that cannot be accommodated, the LOM automatically discards replicas and ROT entries until the required memory is free. Three different space recovery strategies are supported, these are detailed in Chapter 4 and comparative performance results are given in Chapter 6. Replica objects and segments which have been updated by the client are returned to the server responsible for them so that the master copies can be updated. Replica objects which have not been updated are simply deleted from memory, whilst unchanged replica segments require the responsible server to be notified that the replica is no longer held.

### fetch object

To fetch an object the LOM sends a request to the server responsible for the segment to which the object belongs. It then continues to ensure that enough memory remains free to hold the object (estimated by an upper bound), whilst attending to incoming messages. One of the incoming messages will eventually be the replica object or its segment, depending on the group's movement context.

### incoming messages

At the beginning of most major LOM operations, and as part of some of them, the LOM attends to *incoming messages*. Such messages are not part of synchronous operations, like create group or create object, which must block until a reply is received. Incoming messages include:

- group notification — where the LOM is notified of groups created by other LOM instances

- receipts for returned objects and segments — discussed in Chapter 4

- replicas arriving in response to fetch requests

- replicas arriving in response to server generated prefetch requests — see Chapter 4

Replicas arriving in response to fetch requests are placed in memory and their ROT entry notified of their arrival (ROT entries are created prior to a fetch request being generated). In the case of a replica segment, new ROT entries are created as necessary for object replicas present in the segment but not found in the ROT, and a notification of acceptance is forwarded to the server responsible.

### 3.4.3 Global Object Manager (GOM)

The GOM is the layer responsible for server communication with clients and other servers. It implements the operations requested of it in terms of the primitives provided by the storage layer. Unlike the LOM, in which each instance is independent of other instances, the GOM is not a particular instance executing on a server, but should be considered as the collection of all instances executing. The individual GOM instances cooperate to varying degrees on different tasks. One instance is known as the *main server* and is responsible for tasks which are not efficiently distributable. Each group has a *base server* which coordinates the activities of the servers with regard to that group.

The operations which can be requested of the GOM are:

**create group**

In response to this request, the main server passes the parameters defining the group to the associated storage layer operation which creates the group in the group table and returns the entry to the requesting LOM. The new entry is then sent to the requesting client and also broadcast to all clients and servers. This last step is necessary so that other processes can correctly handle objects belonging to that group when they are encountered.

**create object**

An object creation request can only be received by the base server for the group in which the object is to be created. The base server then attempts to find a server for the group which can satisfy this request. This is done by passing the request around the group's servers until it either returns to the base server or is satisfied. To satisfy the request a server must have enough free space in an existing segment of the group to create the object, and if the movement context is by segment, no other client must hold a replica of this segment[6]. If the request returns to the base server, no existing segment can accommodate the request. In this case a server is nominated (this is done in cyclic order) to create a new segment and satisfy the request. The nominated server does so, and replies directly to the requesting client with a newly allocated object identifier.

**return objects**

Returned objects can take three forms — complete objects, complete segments, and notifications of segment deletion. Complete objects are simply copied over their master copies. For complete segments, only those objects which have been updated by the client are copied back. This is necessary to maintain consistency under concurrent access. For both complete objects and segments, a receipt is immediately sent to the client, and timestamps are incremented. In dealing with a notification the server simply deletes the client from the set of clients known to be holding a replica.

**fetch object**

To satisfy a fetch request the server extracts the segment and group information from the object identifier and looks up the appropriate segment using the group table. A replica object or segment, depending on the movement context, is then sent to the requesting LOM. If the group of the object requested has prefetching to a non-zero depth, prefetch requests will be generated for each object referenced by the fetched object. If the object to be prefetched resides on a different server, then a prefetch request message is sent to that server. If it resides on this server, it may be handled either as an explicit message or placed into the prefetch queue.

---

[6]Whilst this second condition may be restrictive for some concurrent applications, the nature of the sample applications is such that it should not have adversely affected results for the movement context experiments (Chapter 7). It is believed that further work on the handling of segments will yield several improvements, including the lifting of this restriction.

**name object**

This operation is only valid on the main server, and registers a name/object identifier association with the mapping maintained in the storage layer.

**object named**

This operation is only valid on the main server. It requests, from the storage layer, the object identifier associated with the name supplied. This identifier is then passed back to the LOM.

**accept segment**

This notification indicates that a client has accepted a replica of a segment, and is maintaining it in local memory. The client is added to the set of those known to hold replicas of the segment.

**Wait**

Associated with each object is a binary semaphore and a queue of those clients which are blocked on a Wait for that semaphore. Upon receiving a Wait request, the server adds the client's identifier and the timestamp of any replica of the object that the client has, to the end of the queue of clients waiting. A *notification* function is then called which, if the semaphore is non-zero and the queue non-empty, decrements the semaphore and removes the client and timestamp from the front of the queue and notifies the client. A replica object or segment is forwarded to the client by the notification function if, and only if, the replica held by it is outdated relative to the object or segment's current timestamp.

**Signal**

This operation increments the semaphore by one and calls the *notification function* described above.

**synchronise**

Each server keeps a count of the number of synchronisation requests received so far. When this count equals the number of clients, it broadcasts an acknowledgement to all clients. Each client blocks until it receives an acknowledgement from each server. This protocol ensures that all messages sent from any client to any server prior to the client synchronising, are attended to by the server prior to *any* client returning from the synchronisation.

**prefetch request**

Prefetch requests are handled in one of two manners. If *high priority* prefetching is being done, the prefetch message is handled as if it were a normal fetch request, except that the degree of subsequent prefetching from the object is decremented by one. *Low priority* prefetching results in prefetch messages being placed on the prefetch queue which is attended to only when no messages are queued. Requests taken from this queue are then handled in the same manner as high priority requests.

In addition to servicing message requests, the GOM can detect when no messages are waiting and proceed to do other work. At present this is only utilised by servicing the prefetch queue under the low priority prefetch policy. However, it may be useful in future for implementing policies such as background garbage collection, dynamic load balancing, or migrating objects to stable storage.

### 3.4.4 Storage Layer (SL)

The storage layer creates and manages the storage units (segments) in which the heap objects reside. It provides the GOM with access to these segments on an "as needed" basis for object fetches and updates. Each storage layer generates a unique stream of object identifiers, one of which is utilised during each successful object creation operation. The storage layer of the main server is also responsible for maintaining the system wide mapping from names to object references. The operations provided by the storage layer are outlined below.

**name object**

This operation is only valid on the main server, and registers a name/object identifier association in the mapping maintained. Any previous mapping involving that name is overwritten.

**object named**

This operation is only valid on the main server. It returns the object identifier associated with the name supplied, or returns NULL_OID if there is no association.

**create group**

To perform this operation the storage layer creates a new entry in the group table, and allocates a group identifier and a base server. The storage layer keeps track of which server should be next allocated as a base by adding the width of the group just created to the base of that group, and taking the result modulo the total number of servers. The new entry in the table is then returned to the GOM.

**new segment**

This operation simply creates a new, empty segment for the specified group.

**create object**

Given the group, the requesting client, and parameters describing the size of the requested object this operation attempts to create the object in one of the existing segments managed by this server for the given group. If successful it returns the object identifier for the object to the GOM, if not NULL_OID is returned.

## Segment Operations

Segment instances are managed primarily by the SL, although migration by segment does allow segments to also exist within the LOM. The following operations are available as methods on the segment data type:

**create object**

If there is sufficient space in the segment, the object is created there and the next object identifier is generated and allocated to it. If successful, the object identifier is returned, otherwise NULL_OID is returned. Additionally, if successful and the movement context of the group is by segment, then the segment's timestamp is incremented.

**copy object**

This operation gives access to the physical memory in which the specified object in the segment is stored, and facilitates the GOM's sending of replicas.

**return object**

This operation replaces the master copy of the specified object with a replica returned from a client, and increments the object's timestamp.

**copy segment**

This operation gives access to the physical memory in which the segment is stored, and facilitates the GOM's sending of replicas.

**return segment**

This operation updates the master copies of objects in the segment from copies in a replica segment returned from a client, and increments both the segment's timestamp and that of the object. Only objects updated by the client have their master copy changed.

**discarded segment**

This operation removes a specified client from the set of those who hold a replica of the segment.

**accept segment**

Given a client, this operation inserts it into the set of those known to hold a replica of the segment.

# Chapter 4

# Data Structures & Implementation

In this chapter, the main data structures of the distributed object server are described in the context of the software layers in which they are used. This is followed by a description of the algorithms of each of the major system operations.

The object server system is implemented in C++, with each of the software layers being an instance of a C++ object. Operations provided by these layers are methods invoked upon the object instances. In the case of the GOM, a separate server function receives messages and invokes the appropriate methods. Each of the layers contains and works with data structures. The basic data structures which are common to most of the layers are described first, then the significant structures in each layer are detailed.

## 4.1 Primitive Data Structures

### group identifier

Group identifiers are generated by the main server as new groups are created. They are 16-bit values in which all combinations are legal. If the 16 bits are considered as an unsigned integer the values are generated from zero upwards. The value with all bits on indicates the constant NULL_GID referring to the null or nonexistent group. For ease of use the 16-bit value is often represented in 32-bit form with sign extension.

**segment identifier**

Segment identifiers are generated by the servers as they create new segments in a group. Each server responsible for a group knows the width of the group, and its own rank within that group (distance from the base server for the group). A stream of unique segment identifiers is generated for a group by each server. They are 16-bit values in which all combinations are legal. The value with all bits on indicates the constant NULL_SID referring the null or nonexistent segment. As with group identifiers, the segment identifier is often represented in 32-bit form.

**location**

A location is a composite structure consisting of a group identifier and a segment identifier. Each location is unique in that it identifies a single segment managed by a single server. Each object has a location, and many objects may have the same location. Both identifiers stored in a location are in 16-bit form, hence a location occupies only 32-bits.

### 4.1.1 Object Identifiers (OID)

An object identifier is a 64-bit composite structure consisting of a location and a unique object identifier. It not only identifies the object, but also provides the means of determining its server and segment.

**unique object identifier**

A unique object identifier is a 32-bit value in which the low order bit is always on. All combinations for the other bits are legal, with the value in which only the two lowest order bits are on being the constant NULL_ID referring to the null or nonexistent identifier. Each server generates its own stream of unique identifiers. These are used to distinguish between objects within segments, but are capable of uniquely identifying objects across the entire space.

### 4.1.2 Object References

It is the object reference data type which is used by client programs to specify objects for access. Object references can exist within the client program's transient address space, and also within the reference part of objects created in the shared heap. They are similar to a C union in that they consist of either an OID or an index into the ROT, these are known as the *unswizzled* and *swizzled* forms respectively. Both forms occupy 64-bits, but can be distinguished by checking the low order bit of the first 32-bit word. The ROT index is a pointer and so will always have a zero in this position, whilst the unique object identifier which forms part of the OID will always have a one. Since an index into one client's ROT is not valid beyond that client, all references in objects being returned to servers by a client must be unswizzled.

### 4.1.3 Segments

The segment is the system's unit of storage in which objects are created and master copies maintained. They are used by the storage layer, and also in the LOM for those groups whose movement context is by segment.

A segment consists of two fixed size structures and a number of attributes. The structures are a *data area* and a *directory*. Each of the structures places a limit on how many objects can be stored in a given segment. The total size of all objects must not exceed the capacity of the data area, and the number of objects cannot exceed the number of entries in the directory.

The attributes associated with each segment fall into three groups, those used by both the SL and LOM, and those used by each alone. The common attributes are:

- group and segment identifiers

- amount of free space in data area

- a timestamp

- number of objects in the segment

The SL alone uses an associated list of clients holding a replica, whilst the LOM uses the following:

- a "dirty flag" indicating that at least one object in the segment has been updated

- a "protection flag" set by the LOM under certain circumstances to discourage the discarding of this segment during space recovery operations

- a reference count of the number of swizzled object references to objects in the segment — used by one of the space recovery strategies

**data area**

The data area consists of a block of memory in which objects are stored. The data and reference parts of each object are stored contiguously. This enables a single offset and the size of each part, which is given in the directory entry, to completely define the memory occupied by a given object. Free space in the data area is "owned" by directory entries having NULL_OIDs. Since object deletion has not been implemented, no fragmentation of free space occurs. There is however, no reason why coalescing of free space should not occur, and in fact some code to do this is already in place.

**directory**

The directory contains a fixed number of entries, each one detailing the space in the data area occupied by an object, along with the object's OID, a timestamp, and semaphore related information. Directory entries are kept in sorted order according to the unique object identifier part of the OID, and a binary search is done to locate the entry for a given OID when needed.

## 4.2 Storage Layer

The main structures of the storage layer (SL) are the group table and named object list. In addition, the SL also maintains a counter from which the next unique object identifier to be allocated is generated, and the main server keeps track of the next base server to be allocated to a new group.

### 4.2.1 Group Table

The group table stores information describing each of the groups created by the main server at the request of clients. It is implemented as an array of group entries, with the group identifiers acting as indexes into the array. Each group entry has the following attributes:

- group identifier (NULL_GID indicates the entry is unused)

- base server

- width

- prefetch depth

- movement context

- counter from which the next segment identifier to be allocated by this server is generated

- linked list segments (or replica segments) belonging to this group which are in local memory

Note that each LOM and SL instance has a group table, and that notification of newly created groups is broadcast to all processes. This broadcast is necessary since any process, client or server, may at some stage need information regarding any existing group. If a client or server does at some stage require information regarding a group which is not yet in its group table, it blocks until the necessary notification arrives. This assumes reliable communication and non-forgeability of group identifiers.

Deletion from the group table is not supported. The semantics of such an operation are not intuitive in the face of possibly existing objects and references to objects in a group. The existence of unaccounted group identifiers in the transient memory of client programs also presents problems.

The mapping from locations to servers is defined below. It assumes that the servers are labelled with integers beginning at zero and increasing by one, and that the servers allocated to a group cover a continuous range modulo the total number of servers.

$$server\_rank = ((seg\_id \bmod group\_width) + group\_base) \bmod num\_servers \tag{4.1}$$

### 4.2.2 Named Object List

This is simply a list of name-OID pairings. It is implemented as a linked list and can be traversed searching for the OID matching a supplied name. If no matching name is found, NULL_OID is returned. Many OIDs may be associated with any name, but only the most recent pairing will be in effect at any time. Names are any valid NULL terminated character string.

## 4.3 Global Object Manager

The global object manager (GOM) instances contain only two attributes and one data structure. The attributes are a synchronisation counter and a client closure counter. The synchronisation counter indicates how many clients have so far requested a synchronisation with this server. When all clients have done so the server acknowledges each one's request and resets the counter to zero. As each client process terminates, it returns all updated replicas to the servers and notifies each server that it is "closing". When a server has received closure notices from all clients it may safely terminate.

### 4.3.1 Prefetch Queue

The prefetch queue is where prefetch requests to be acted upon by this server are stored under the low priority prefetch scheme. With high priority prefetch the prefetch requests are sent as messages by the server to itself. A prefetch request is represented in the queue by an OID-client pair. The following queue operations are supported:

- determine number of entries

- take entry from front of queue

- add new entry to front of queue

- add new entry to end of queue

- remove all entries from queue

## 4.4 Local Object Manager

The local object manager (LOM) contains two attributes and several data structures. One attribute defines which space recovery strategy is presently selected — this can be toggled via an API function call. The other determines which swizzling policy to use. Swizzling policies are not within the scope of this investigation, and the simple policy of swizzle upon demand was used throughout.

The data structures include a group table (described above in §4.2.1), the resident object table, and two receipt pending lists.

### 4.4.1 Resident Object Table (ROT)

The ROT is the most used data structure in the entire system. Every object access utilises it, each object reference used searches it at least once, and each of the major LOM management tasks modifies it. With every resident object replica having an entry, and the local memory capable of holding tens of thousands of replicas, the ROT must efficiently store and facilitate the manipulation of a large number of entries. Additionally, two of the three space recovery schemes implemented require information about each resident object replica and it is convenient for the ROT to gather and make available this information.

It is the structure of the ROT which has the greatest impact on the efficiency of operations which manipulate it. The information content of individual entries does not affect the operational efficiency other than indirectly through spatial efficiency. For this reason the ROT structure and entries were implemented separately.

**ROT Structure**

The ROT structure must be operationally efficient for a large number of entries. The basic operations which must be provided are insertion, deletion, and search for key. Least Recently Used (LRU) information which the space recovery strategies need on a per object basis, is most conveniently represented structurally as this can facilitate a traversal of the entries from LRU to MRU (most recently used).

**Doubly-Linked Self-Organising List** — the initial implementation of the ROT structure was a doubly-linked, self-organising list. This was maintained in MRU order, the order being updated with each object access through the ROT. Insertions went to the head of the list, and deletions and searches required searching the entire list. Traversal in either LRU or MRU order was easily accomplished due to the double linkage.

This structure was found to be adequate for cases where the ROT numbered only a few hundred entries. Not surprisingly it scaled appallingly as the entries grew to be numbered in the tens of thousands.

**Self-Organising RED-BLACK Tree** — using the fact that the unique object identifier portion of each OID can provide an ordering, a binary tree was considered for a second implementation of the structure. It soon became apparent that a normal binary tree would however give far from optimal performance. As a process creates the objects it requires for a computation, a long sequence of ordered OIDs may be inserted into the structure. This would result in a skewing of the tree, and in the worst case yield performance the same as the list structure tried previously. To overcome this problem the RED-BLACK tree [11] form of the binary tree was implemented. The RED-BLACK tree has similar properties to the AVL tree [22], and guarantees that the tree is always nearly balanced. Insertions and deletions trigger reorganisations if needed to maintain the properties. Applying the tree structure significantly improved the time complexity of the search and delete operations (from $O(n)$ to $O(\log n)$), at the expense of the insertion operation ($O(1)$ to $O(\log n)$). When dealing with large $n$, and searches that are more frequent than insertions, the savings are substantial.

The tree structure is not useful for maintaining LRU information or traversing the structure in LRU or MRU order. For this reason, the nodes of the tree are also threaded in a doubly-linked self-organising list as per the initial implementation. Hence, there are two views of the ROT structure. One is the tree view used for fast searching, and the other is that of a list in MRU order which can be traversed in either direction. Figure 4.1 shows both views of a small ROT structure.

**ROT Entry**

The information contained in each ROT entry is used by the system to determine whether the associated object replica is still in local memory, and if so, how to access the replica. Other information is provided to facilitate the space recovery task. The attributes maintained in each ROT entry are:

**OID**  of the associated object.

**"dirty flag"**  indicating whether the local replica has been modified.

**"protection flag"**  set by the LOM on occasions to discourage the discarding of the associated object, and forbid the deletion of this entry in any case.

**reference count**  of the number of swizzled object references to this entry.

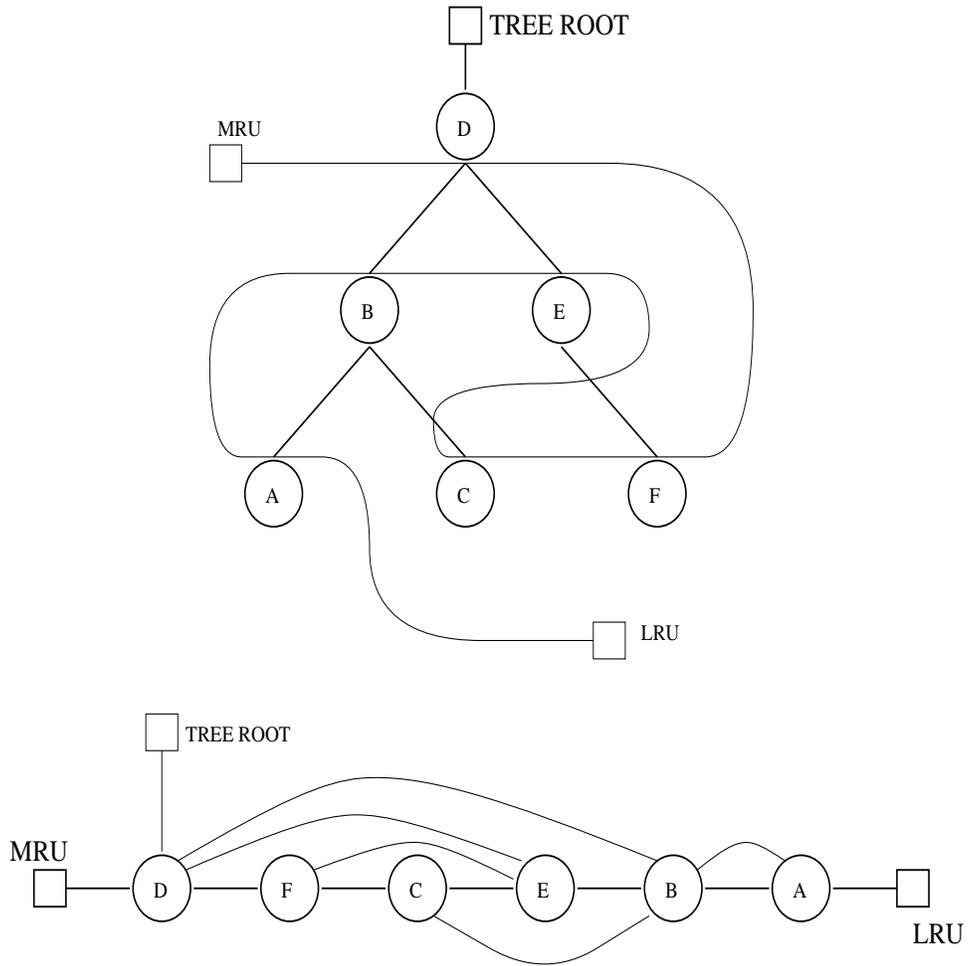

Figure 4.1: Two views of the same ROT structure

**segment pointer**  indicating the local replica segment in which the object resides, NULL if none.

**object wrapper pointer**  indicating the object wrapper to be used to access the object replica, NULL if there is no replica present. A wrapper is present if and only if an object replica is present; this includes object replicas which are part of a replica segment. Object wrappers are discussed further below.

### Object Wrapper

Each object replica (including those in replica segments) in local memory is accessed through a wrapper object. The wrapper prevents direct access to the physical memory in which the replica is stored. The need for indirect access arises from the possibility of any operation on a store object triggering a space recovery task. Depending on the strategy used to recover space, some or all of the resident replica's will be removed from memory. If one of the removed replicas is subsequently accessed the system needs to be able to intercept the access, redirecting it to a subsequently fetched replica for the same object. The wrapper also includes information which it is neither necessary nor efficient to maintain in the ROT entry. Attributes such as the size of the object's data and reference parts, and time stamp, are not useful if a replica is not actually present[1].

Figure 4.2 shows the relationship between a swizzled object reference and the physical memory occupied by a replica in local memory.

### 4.4.2  Receipt Pending Lists

There are two receipt pending lists in the LOM, one for objects and one for segments. Whenever an object or segment is returned to a server, an entry for it is added to the end of the appropriate receipt pending queue. Upon receiving the returned item the server issues a receipt to the client. When the receipt arrives and is attended to by the client (as one of the incoming messages mentioned in §3.4.2) the matching entry is removed from the receipt list. Since pending receipts are added to the end of the list, the actual receipts will arrive in list order.  The motivation for maintaining receipt pending lists is given in the discussion of prefetching in §4.7.

---

[1] Space recovery removes replicas but cannot necessarily remove their ROT entries (object references may still contain a pointer to the entry), hence storing these attributes in the ROT entry would waste space.

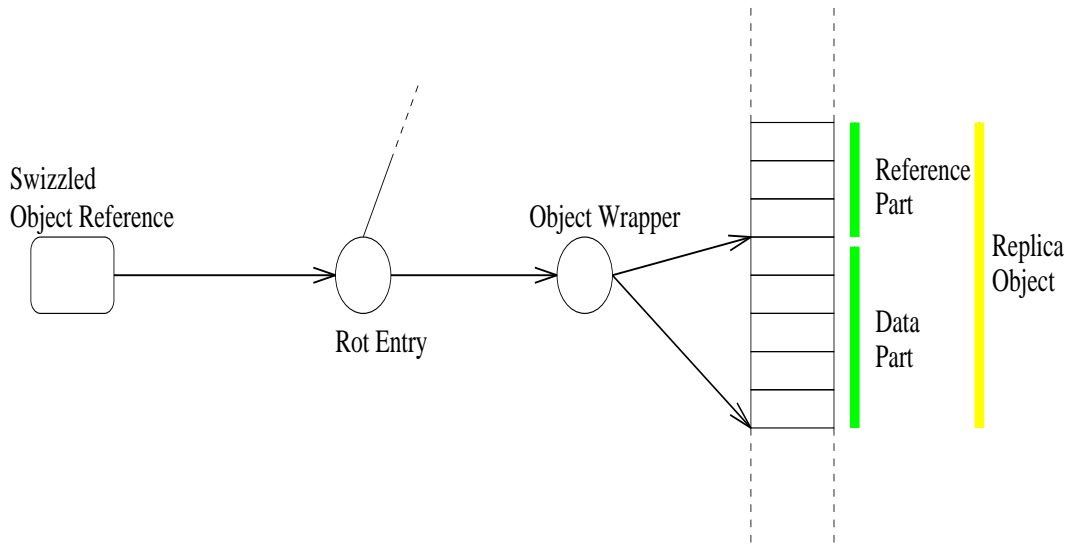

Figure 4.2:  Relationship between a swizzled reference and the referenced object

## 4.5   API

Unlike the other software layers the API is not implemented as a single C++ object.  It does include an API object which provides some of the services, but the object manipulation operations — read, write, Wait, Signal, reference assignment, and indexing — are methods invoked upon object references and object reference wrappers.

### 4.5.1   Reference Wrappers

As noted previously, direct access to the physical memory of replica objects by client programs is dangerous.  This problem has been solved by making object references access replicas through the ROT and object wrapper.  A related problem arises when dealing with references obtained by indexing. Consider the following C++ assignment statement:

```
q[n] = some_function(x)
```

The order of operations imposed by most compilers on this statement is as follows:

1. evaluate the l-value `q[n]`

2. evaluate the r-value `some_function(x)`

3. copy the r-value to the l-value

This ordering is acceptable for conventional data objects in the transient heap, because step 2 cannot affect the result of step 1[2]. If `q` is an object reference, then `q[n]` is an object reference obtained by indexing, and as an l-value is a physical memory address. Subsequently evaluating `some_function(x)`, may involve operations which trigger space recovery and remove the replica object referenced by `q` from memory. The l-value obtained in step 1 thus becomes invalid, the r-value being copied to where `q[n]` was. In such a case, we need to fetch `q` and compute the l-value only after the r-value has been evaluated.

The solution implemented was to have indexed references evaluate not directly to the intended reference, but to a reference wrapper object. The reference wrapper contains only two attributes, the object reference initially indexed, and the index value used. The calculation of the l-value for the actual reference to be assigned is then delayed until step 3 where an overloaded assignment operator performs the necessary operations. The l-value of the reference wrapper is calculated in step 1 and since the wrapper is constructed just prior to the evaluation of its l-value, side-effects on `q` or `n` in calculating the r-value will play no part in the actual assignment. A revised order of operations for the troublesome assignment is thus:

1. evaluate the l-value `q[n]`

    (a) construct the reference wrapper

    (b) evaluate the l-value of the reference wrapper

2. evaluate the r-value `some_function(x)`

3. copy the r-value to the l-value

    (a) evaluate the l-value of the wrapper's indexed reference

    (b) copy the r-value to the l-value

---

[2]This ignores the possibility of `n` or `q` changing as a side-effect of `some_function(x)`. It is assumed that the values prior to step 1 are those intended.

## 4.6 Space Recovery

It is often the case that an application wishes to work with a set of objects whose total size exceeds the capacity of the local memory allocated to caching replicas. The ratio of the size of the set of objects accessed by an application to the size of the client's replica caching memory is termed the *memory access ratio*. In order to accommodate the extra replicas, it is either necessary to enlarge the replica cache or recover space in the existing cache by removing replicas currently stored there. Enlarging the cache is not a general solution since it is limited by the total amount of memory present; flushing some replicas from the cache is however a general solution capable of simulating an infinite cache at the expense of re-fetching some replicas.

There are many strategies for determining which replicas to discard when the managed area of local memory becomes full. Three have been implemented and are described below. The first discards all replicas indiscriminately, whilst the other two attempt to maintain the working set of replicas by discarding a smaller set based on information in the ROT.

### Targeting

The functions implementing each of the three strategies described below have a common interface. The one parameter required by each is the amount of managed local memory that should be free when the function exits. This is called the *target*. Obviously the target should be at least as large as the memory request which triggered the space recovery operation. This is however not optimal. When a space recovery operation is triggered, the local memory will contain two classes of replicas: those which will be accessed in the near future, and those which will not. The operation will have been triggered because the client program wishes to access (or create) an object for which there is no local replica. Removing only enough replicas to satisfy the immediate request will leave the local memory at capacity, meaning that any subsequent access to a non-resident object will trigger another space recovery operation in addition to the fetch or create operation that would otherwise be required.

In response to this problem, the target amount to be cleared whenever a space recovery operation is triggered is a function of the memory requested. The targeting function implemented (shown in Figure 4.3) is bounded below by a fixed portion of the local memory, and reduces to the identity function when the amount requested is large relative to the local memory. Note that

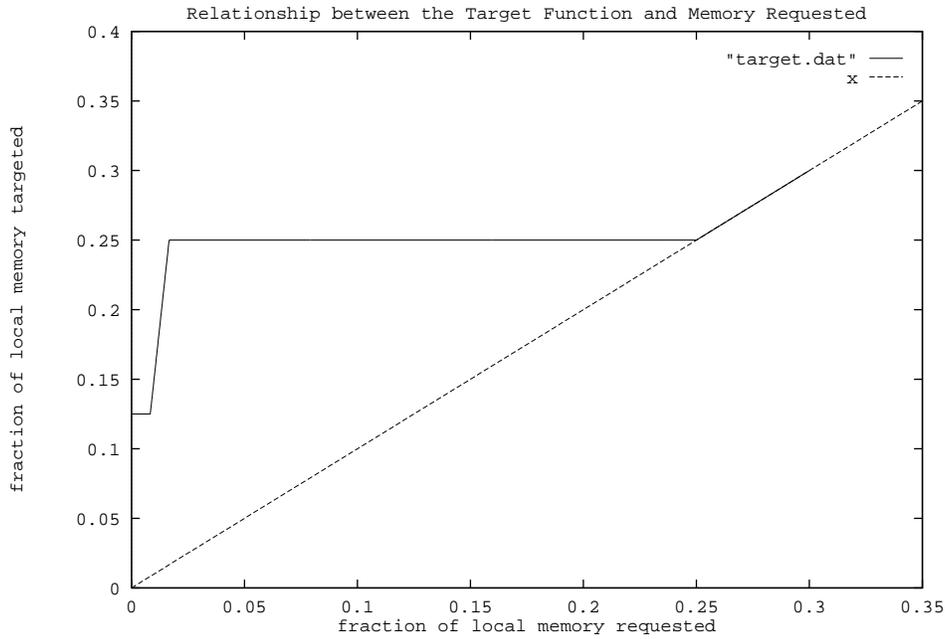

$$target(x) = \max(x, \min(T * HI, \max(T * LO, K * x)))$$

$T$ is total local memory (1 in above graph)
$K > 1$ is a constant multiplier (15)
$HI$ and $LO$ are constant ratios (1/4, 1/8)

Figure 4.3: The Targeting Function

the size of a segment places an effective upper bound on the amount of memory that will be requested at any one time. With a local memory in the order of megabytes and a segment size of 100 000 bytes or less, memory requests would not exceed 10 percent of the memory size, and would usually be less that 5 percent. Hence the greater adaptivity of the targeting function for requests in this range.

### 4.6.1 dumping

This is the simple strategy of removing all replicas from memory. Thus, each space recovery operation necessitates the subsequent initial object access triggering a fetch operation. The value of the target parameter does not affect the behaviour of this strategy in any manner.

This strategy is currently implemented as an invocation of the simple LRU method described below, with total local memory as the target. Whilst this is not an optimised implementation of the strategy, it is believed sufficient to determine the merits of this strategy relative to the other two.

### 4.6.2 simple LRU

This strategy utilises the LRU information maintained by the ROT and discards only enough replicas to meet the target requirement. It traverses the list view of the ROT in LRU order, constructing lists of object and segment replicas to be discarded, and of ROT entries that are candidates for deletion. Deletion candidates are those ROT entries whose associated object is not present, which have a zero reference count and are not protected. When the total size of replicas marked for removal exceeds the target, the ROT traversal halts. The replica lists are then passed to discarding functions, one each for object and segment replicas. These functions unswizzle references contained in the replicas, return them to the servers (or notify the servers) as necessary, and delete them from memory. Receipt pending entries are added to the appropriate lists as needed. The object discarding function buffers messages to the servers until either it completes or 500 000 bytes of messages have been accumulated. The segment function sends each segment immediately it has been packed, in the belief that the server can make progress in the processing of the segment sent whilst the next one is being packed.

Once the replicas have been removed the ROT entries marked for deletion are also removed. The process of selecting entries for deletion prior to removing replicas means that the entries for replicas deleted in the most recent operation are still present. Thus, if the replica is accessed again soon, the two ROT manipulations have been saved.

A notable flaw in this strategy arises when dealing with segments. All objects of a replica segment are represented in the ROT. If one of these objects has not been recently referenced when a space recovery operation is triggered, then it will be near the head of the LRU list and a candidate for removal. Removal of the object, and its associated replica segment, will remove the other more recently used objects of the segment as well. The classified strategy described below takes this and other information into account, but is far more complex as a result.

### 4.6.3 classified

This strategy utilises more of the information available in the ROT to determine the set of replicas to be removed. It traverses the entire ROT in MRU order, allocating each replica object or segment to one of eight classes. Objects, or segments containing objects, which are protected are not included in any class. Ordering within the classes (Figure 4.4) is LRU to MRU, with items inserted in MRU order. The three attributes upon which the classification is based are available from the ROT entry:

- is it a replica object or replica segment?

- has the object been updated (dirty or clean)?

- is the reference count zero (no local refs or local refs)?

During the ROT traversal, a list of ROT entries to be deleted is also constructed. After traversal, the segment classes have duplicates removed, with the entry corresponding to the MRU object of the segment being maintained in the class. Replicas are then discarded from each class in turn, using the order given in Figure 4.4, until the target has been reached. The functions described in §4.6.2 perform the actual removal operations and notify the servers as necessary. Following this, the ROT entries marked for removal are deleted.

The idea behind the classification system is to identify those replicas which are less likely to be accessed in the near future, and are cheap to remove. These are then discarded prior to replicas which are likely to be accessed in the near future and/or are expensive to remove. Replicas whose ROT entry has a non zero reference count are considered as more likely to be referenced in the near future than those with a zero reference count. It is considered that dirty replicas are expensive to remove because they need to be sent back to a server so that the master copy can be updated. Clean replicas, on the other hand, can be removed straight from memory, segments do require server notification but this is relatively cheap and does not even generate a receipt. Based on these assumptions, replicas are classified into

1. no local refs $_\wedge$ clean $_\wedge$ segment

2. no local refs $_\wedge$ clean $_\wedge$ object

3. no local refs $_\wedge$ dirty $_\wedge$ segment

4. no local refs $_\wedge$ dirty $_\wedge$ object

5. local refs $_\wedge$ clean $_\wedge$ object

6. local refs $_\wedge$ clean $_\wedge$ segment

7. local refs $_\wedge$ dirty $_\wedge$ object

8. local refs $_\wedge$ dirty $_\wedge$ segment

Figure 4.4: Space Recovery Classes

the eight classes shown in Figure 4.4, which are discarded from local memory in the order shown. The classes are ordered primarily on the existence of references to the object, then on the basis of whether it is clean or dirty. The third level of ordering, segment and single object, is based purely upon intuition and was not considered in the course of this project.

This strategy does not remove from memory replicas which are marked as protected. It is thus possible that the operation as described can run to completion without achieving the required target. A fall back mechanism has been implemented to avoid this problem. In the case that the strategy fails to achieve its target, a call to the simple LRU strategy is made with the same target.

## 4.7   Prefetching

Many data structures have common patterns of usage; that is, the elements in the structure are often accessed in similar patterns. One of the most common usage patterns is to follow one or more of the pointers contained in a node soon after that node is accessed. For example, a linked list is often traversed in a pattern consisting of examining the value of the current node and then following the pointer to the next node. This pattern occurs as part of most list operations – search, duplicate deletion etc. Similarly, when traversing binary trees it is likely that at least one of the pointers in a node will be followed soon after the node is first examined.

Generalising this common pattern to the arbitrary directed graph formed by the objects stored in the shared heap, improved performance of the fetch operation may be achieved by the prefetching of objects likely to be accessed next. A prefetching method will attempt to fetch the next object from the server while the client is processing a previously fetched object.

The procedure followed when a client attempts to access an object for which it holds no replica is that a request is sent to the GOM, which services it by supplying a copy of the required object. The prefetch mechanism implemented within the global storage manager attempts to preempt some future fetch requests. When a request for an object is satisfied in the normal manner the desired replica is forwarded to the client. Optionally, the server then queues a request on behalf of the client for those objects referred to in the reference part of the requested objected just sent. This occurs transitively to a specified depth, and we refer to this as the *prefetch depth*. Thus, if the client process was to subsequently require any of the objects whose references are contained in the reference part of the requested object, these will have at worst already been queued for sending, and at best have already arrived at the client. This mechanism gives the possibility of reducing both the number of fetch requests and the average time required to satisfy those that do occur.

The clients' pattern and frequency of access determines whether each new access incurs small delay, being met by the prefetch from a previous real request, or whether the server delays induced by serving unnecessary prefetch requests outweigh the benefits to the client.

The client side can suffer from more subtle costs as well. These include quickly filling the local memory necessitating more frequent clearances, and overhead time spent receiving extra objects. At a more general level there is also the problem of generating extra network traffic.

### prefetching priority

Two prefetching mechanisms have been implemented in the GOM. The first, called *high priority* prefetch, sends a message for each prefetch request regardless of which server it is to be serviced by. Thus, a server may well send messages to itself. These messages are received via the normal message queue and are acted upon as soon as received. This equates to a breadth first servicing of prefetch requests, with transitively generated requests being added to the message queue after their parent's siblings. This mechanism gives prefetch requests equal priority with messages arriving from the client processes.

The second mechanism implemented is called *low priority* prefetch. It uses messages only when queueing prefetches on another server. All prefetch requests to be serviced by the server itself are placed in a prefetch queue, as are requests received from other servers. The server attends this queue whenever the message queue is empty, thus giving the prefetch requests lower priority than client generated requests. It is possible to add prefetch requests to the queue in such a way as to generate breadth first or depth first traversal. Both have been implemented and investigated.

### client-side

Dealing with the replicas forwarded to the client by prefetch requests is not a simple task. Firstly, the messages are not expected, and cannot be, since under the low priority mechanism they may not actually be sent. This is overcome through the use of an incoming messages function which is called at the commencement of each major operation and is an integral part of the fetch operation. This function clears the message queue of all incoming messages such as receipts, prefetched replicas, etc. Once a replica is received in this manner, the LOM must decide whether to accept the replica into local memory. A replica is accepted into local memory if and only if each of the following hold:

- there is sufficient free space in local memory without inducing a space recovery operation;

- the received replica is newer (by timestamp) than an existing replica for the same object/segment, or there is no such existing replica; and

- there is not a receipt pending entry for the replica.

The rationale for the first condition is obvious, in that the expense of space recovery operations should not be induced to accommodate replicas which may not be needed. To do so would be to replace replicas which were definitely used with those which only may be required. The second condition specifies that replicas which are out of date or contain no new information should not be accepted. The only manner in which a replica could be newer than one already in local memory is if it has been updated by another client process. These replicas can be safely accepted provided appropriate coherence mechanisms are employed by the client program. When newer replica segments are accepted, only those individual objects within them which have been updated are accepted, and this is done by copying the updated data and reference parts to the physical memory occupied by the existing segment.

Refusing a replica on the basis of a receipt pending is perhaps the most interesting of the conditions. When a replica is returned from the client to a server, the client adds a receipt pending entry to the appropriate list. The server responds with a receipt when it processes the message containing the replica. A replica arriving at the client for which there is a receipt pending entry, was obviously sent from the server prior to the server processing the returned replica and generating a receipt. Hence, the replica which arrived at the client is out of date with respect to the replica the client returned to the server.

## 4.8 Overview

The system has a client-server architecture consisting of four software layers. The server consists of a storage layer (SL) and a management layer (GOM) which services requests from the local object manager layer (LOM) of client processors. Client programs interact with the shared distributed heap through the API layer which consists of methods on both an explicit API object, and upon the user accessible data type known as the object reference.

Figure 4.5 shows the relationship between software layers in the context of the client-server processors.

# ARCHITECTURE OF THE DISTRIBUTED OBJECT SERVER

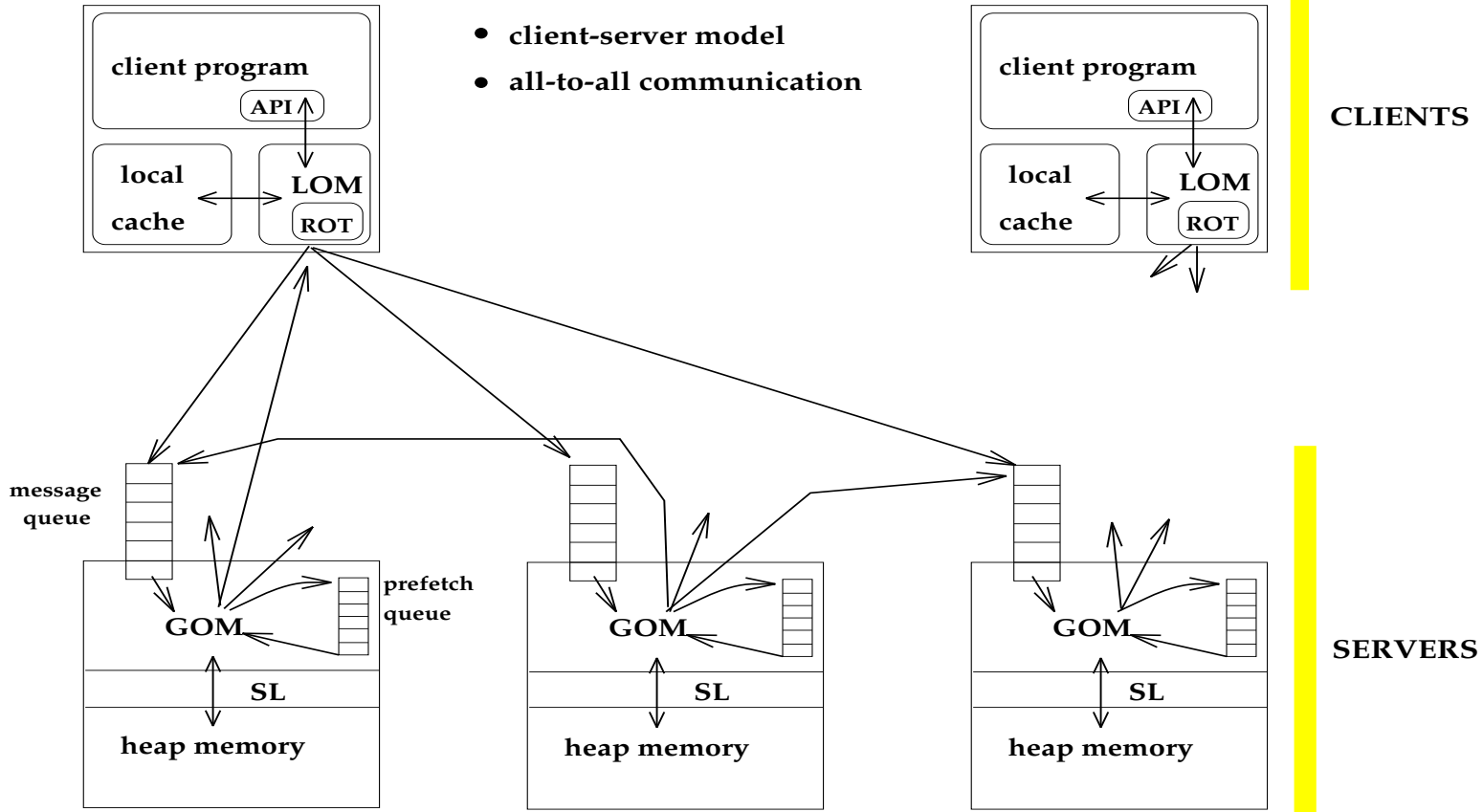

Figure 4.5: System Overview with Software Layers and Message Flows

# Chapter 5

# System Instrumentation & Example Applications

Investigating issues in object store implementation on an MPP machine involves exploring a multidimensional space. Several of the dimensions are quantifiable and easily controlled, these include:

- number of processors

- ratio of clients to servers

- problem size

- object size

- amount of memory

- prefetch depth

Others, such as the pattern of object usage, are implicit in the applications used to test the system and are not readily quantifiable. Further dimensions, execution time and operation counts for example, are quantifiable but dependent upon those already mentioned. This chapter outlines how the distributed object server was instrumented to gather information about the dependent variables. It also details the example applications used to test system performance and why they were chosen.

## 5.1 Instrumentation

The distributed object server constructed is an experimental system used to determine the viability and relativities of various implementation choices. As such, it is necessary that quantitative information regarding the system's performance under a given set of choices be easily obtained and recordable. The system was constructed with integrated instrumentation facilities for both client and server processors. In both instances the major operations of the management layers were instrumented with counters and clocks[1]. The counters reside in instances of special C++ objects within the management layers. A function within each management layer returns a copy of this object with attributes also being set to the current values of the various clocks.

### 5.1.1 client

Instrumentation of a client process is available at any time during execution via a call to the `current_stats()` function which returns a C++ object containing the instrumentation information. Operations valid on this object type include:

**assignment** — allowing intermediate instrumentations to be recorded in variables

**display** — sending all attributes to `stdout`

**subtraction & addition** — returns an object in which the operation has been applied to those attributes (mainly counters and total times) for which the operation has natural semantics, and in which the remaining attributes have the value of the corresponding attribute in the first operand

Values available through the instrumentation facility include counts, total times, maximum, minimum, and average times for each of the major operations, total execution time, and other miscellaneous counters.

---

[1]The clocking module used was a modification of the one given by Spuler in [34].

### 5.1.2 server

Whilst the server instrumentation mechanism works in manner similar to that of the clients', the information is available only when sent to `stdout` just prior to the server exiting. In addition to the instrumentation of the major operations, the server also reports its identity, the number of messages it received, and its idle time.

## 5.2 Example Applications

The example applications were selected with respect to two conditions. They had to make extensive use of distributed heap objects, and as a group have a variety of object access patterns. To facilitate experiments with concurrency control measures it was also required that some be distributed applications.

Four applications were chosen, these being a binary tree construction and search, two $N$-body simulations, and a plucked string simulation. The $N$-body simulations having radically different data structures and usage patterns. The plucked string simulation and one of the $N$-body simulations are distributed applications.

### 5.2.1 Binary Tree Search

Distributed tree structures are useful in many applications, database indexes and $N$-body simulations (see §5.2.3) are but two examples. The ability to handle such structures efficiently is a necessity for any object server.

This application was included primarily as a basic test of object migration schemes. It consists of inserting a fixed number of randomly generated keys into a binary tree. This is followed by a large number of searches of the tree for a further set of randomly generated keys, some of which were not present. Since binary tree insertion is primarily a search, the dominant access pattern is a series of random descents into the tree. An interesting feature of this application is that order in which objects (nodes) are created bears no relation to the order in which they will be required in later searches. Also of significance is the low branching factor of the tree structure. Both of these features were hypothesised to be potentially significant for the performance of migration schemes.

### 5.2.2    $N^2$ $N$-Body Simulation

The $N$-body problem is a well known problem in physics. A physical system consisting of a large number of particles freely interacting with each other is simulated by calculating forces and applying approximate accelerations for small time increments, updating the particles' positions, and repeating this process for each time step. In the initial phase a large number of particles (representing stars, for instance) are generated with a range of masses and positions in 3 dimensions. The main loop repeats force calculations, by summing the force on each particle due to every other particle, followed by updating the position of every particle. The force summation is (naively) of order $N^2$ in the number of particles. This application is an implementation of the naive solution.

Each client process works independently to solve the same problem, and does not cooperate with or in any way access objects of other client processes. This implementation was utilised in gauging the performance of aspects of the distributed server which were not directly concerned with maintaining coherence. The data structure created in the heap by each instance of this application consists of two groups of objects, each representing a group of particles, and an *index* object to each of these groups — see Figure 5.2.2. The index object is a heap object containing only references to the particle objects in its group. The particle objects for a group are thus accessed by indexing into the index object, and the client process needs only maintain a single reference to the index object permanently in its memory. During any one time step, one of the two groups of particle objects represents the current position and velocity of the particles, and the other is an update group in which the updated particle positions and velocities are recorded. On the subsequent time step the roles of the two groups are reversed.

The access pattern during time step simulation is significantly different from that of the setup phase in this application. The setup phase sees each of the groups being added to in turn, whereas the simulation of a time step consists of $N$ (being the number of bodies simulated) traversals of the entire current position group (hence $N^2$ operations) interleaved with updates of different objects in the update group.

This application was included in the suite of test programs because it was believed that the dominant access pattern of repeatedly cycling through a set of objects not directly referencing one another would well exercise the space recovery strategies tested. It was also thought that the spatial locality derived from creating a set of particle objects in the order in which they would be traversed may yield good performance when the movement context was by segment.

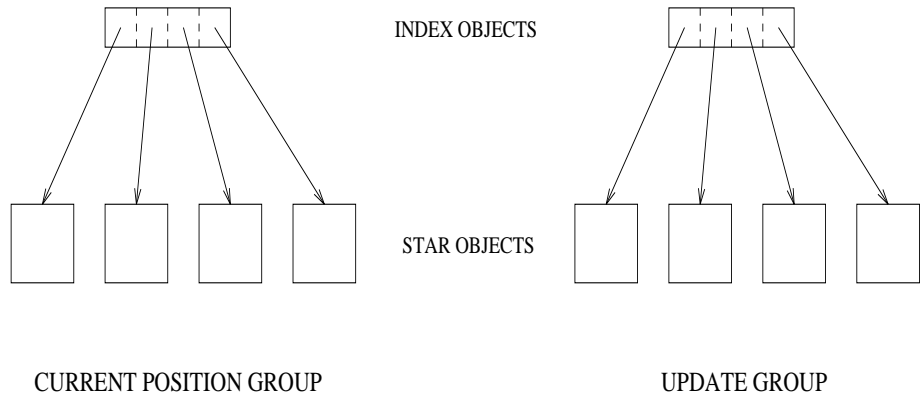

Figure 5.1: Data structures constructed in the distributed shared heap by the $N^2 N$-body simulation

### 5.2.3 Oct-Tree $N$-Body Simulation

This second $N$-body simulation constructs and traverses a very different heap data structure. For a large number of bodies[2] a substantial decrease in computation time can be made by approximating the effect of a group of particles that are relatively far away from a particle of interest. The effect of the group is approximated by the effect of a single particle with the sum of their masses placed at the group's centre of mass. The approximation can be efficiently implemented with the aid of a spatial oct-tree structure, representing combined masses for each group at the internal nodes of the tree. In practice the use of this approximation reduces the computation time to approximately $O(N \log N)$ [39]. A spatial oct-tree has the general form of a node being an approximation for the masses in a volume of space. If this node represents more than a single mass, it has eight children each being the approximation for one octant of their parent's volume.

The programming of the oct-tree based simulation represents a challenging task on distributed memory machines if the programmer needs to manage distribution explicitly. An alternative is the creation of a globally linked data structure in the shared heap. All client processes are then able to transparently access those portions of the oct-tree required for their calculations and update nodes for which they are responsible.

---

[2]The exact number at which savings relative to $N^2$ implementations becomes apparent is implementation and hardware dependent. Experience has shown that for hand crafted code on a 128 cell AP1000, $N^2$ methods are superior for up to 10 000 particles which was the limit of testing.

For this example, the application has two phases: a structure generation phase, involving tree and list building by one processor, and a calculation phase. During the latter phase multiple processors each assume responsibility for a portion of the particle list. They proceed, in parallel, to traverse the approximation tree in order to calculate the forces acting upon each of the particles in their list, synchronising at the completion of each time step.

Both parallel (described above) and serial versions of this application were constructed. The serial version has individual processors constructing independent oct-trees within the heap, and traversing just this tree to apply approximations to their particles.

Two distinct phases, each with a separate access pattern, are apparent in this application. Tree construction consists of a series of searches through the existing oct-tree and the insertion of new leaf nodes as required. The simulation phase consists of a force accumulation pass, followed by an update pass. The accumulation pass sees the client total the forces acting upon each particle it is to update, with the accumulated forces being applied during the update pass. The depth of traversal varies across the tree during the accumulation pass, and particles are accessed directly during the update pass. With particles inserted in random order, spatial locality is unlikely to match temporal locality. It was hypothesised that this would impact upon context movement by segment, whilst the high branching factor of the oct-tree would adversely affect the prefetching strategy.

### 5.2.4 PLUCK String Simulation

The PLUCK problem is to model a taut string that has been plucked. The important features of the problem are;

- the string is stretched between two immovable endpoints.

- the string has some mass per unit length.

- the string is springy, any displacement from the rest position will induce a restoring force $F = -K\Delta x$ (*Hooke's law*: $F$ denotes the restoring force, $K$ the constant of proportionality, and $\Delta x$ the displacement).

- damping is not considered, so the string will oscillate forever.

- a pluck is a displacement of some size at some point along the string

The problem can be solved analytically as a hyperbolic partial differential equation. For the purposes of this exercise however, an approximation by finite element simulation is considered. The string is considered to be a finite collection of rigid elements, each individually reacting to their environment (their displacement relative to their neighbouring elements). Time is stepped forward discretely, with the position of each element being updated at each step.

Two solutions based upon a data decomposition of the problem have been implemented. One uses the object server described in the last two chapters, the other using explicit message passing routines provided by the AP1000's CELLOS operating system. The decomposition and simulation mechanism used in each program is identical, only the information passing and synchronisation methods differ.

A simple block data decomposition, as shown in Figure 5.2, was used. Each process working on the problem is responsible for advancing a contiguous block of elements throughout the simulation. The dependency of each rigid string element on the position of itself and its neighbours in the previous time step introduces the interprocess data dependence shown. This requires a synchronisation and information flow between adjacent processes at each time step. In a manner similar to the $N^2$ $N$-body simulation above, each process maintains two copies of the string elements it is simulating. One copy records the string position at the last time step, the other is an update copy used to record the string position being calculated in the current time step.

The object server version of the application views each rigid element as an object. The binary semaphores attached to the objects representing the ends of the individual client intervals are used to guarantee synchronisation and information flow between the clients. In an initialisation phase all clients create the objects for their interval of the string and register their endpoints in a globally accessible object. Any reasonable use of the Wait and Signal operations on the appropriate endpoint objects will guarantee exclusive access to a client, but ensuring synchronisation without generating deadlock requires careful ordering. The pseudo-code of Figure 5.3 indicates the ordering required.

The message passing version of this application represents each element as an array entry. Endpoint values are exchanged as messages with adjacent processors. The use of blocking receives ensures synchronisation, and simply ordering the sends prior to the receives avoids deadlock. This implementation uses only four operations per time step to achieve the information passing and synchronisation done in eight operations by the object server version.

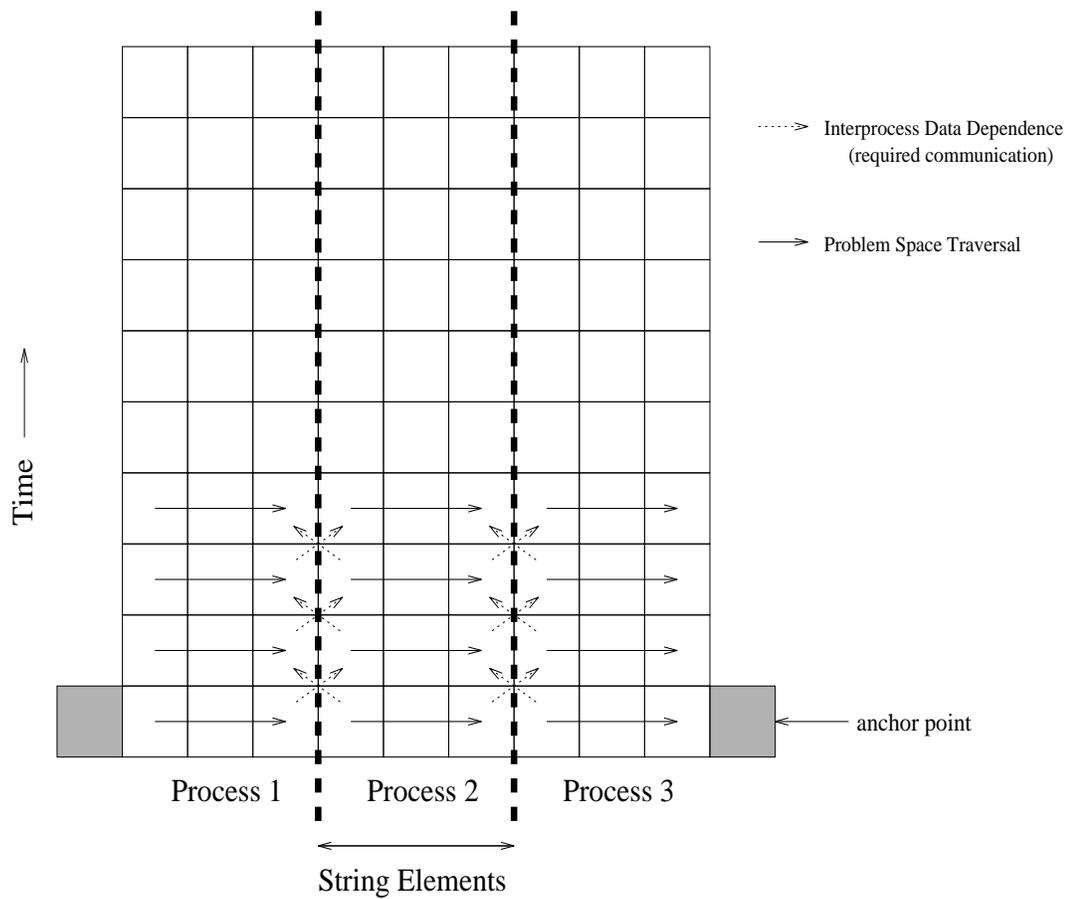

Figure 5.2: Parallel Data Dependence of PLUCK String Simulation

This application parallelises well, and has a predictable low degree of concurrent object accessing. With at most two processes needing to access any one object, and a message passing version for qualitative and quantitative comparison, this application is a good basic exercise of the coherence maintenance mechanism of the distributed object server.

```
local_elements_left_end[prev].Wait();
local_elements_right_end[prev].Wait();

for each time step do

        swap curr and prev

        left_processor[prev].Wait();
        right_processor[prev].Wait();

        for each local rigid element do
                calculate current position

        left_processor[prev].Signal();
        right_processor[prev].Signal();

        local_elements_left_end[prev].Wait();
        local_elements_right_end[prev].Wait();

        local_elements_left_end[curr].Signal();
        local_elements_right_end[curr].Signal();
```

Figure 5.3: Pseudo-Code of PLUCK String Simulation

# Chapter 6

# General System Performance

This chapter describes a series of initial experiments which were performed to determine the general characteristics of the implementation. The performance of all three space recovery strategies is considered under the variation of the *memory access ratio*, the number of servers, and the number of clients.

## 6.1  Memory Access Ratio

Two initial experiments were performed to determine the performance of the system over a range of memory access ratios. Both serial versions of the $N$-body simulation were run for varying degrees of memory access ratio. The client cache used in both cases was 3MB, and the problem sizes were selected so that a similar number of heap objects was manipulated in each case.

Since the number of operations performed by each of the simulations varies with the problem size, it was decided not to generate the increasing memory access ratio by increasing the problem size. The approach taken was to keep the problem size and client memory constant for each simulation, varying instead the size of the objects being manipulated. Each object consisted of the space required to contain the necessary information, and an additional amount of unused space which could be varied. The effect of varying this unused space across all the objects manipulated by the simulation was to change the memory access ratio. The unused space in an object is termed the *object fill* and is considered a measure of object size, since it does in general dominate the information content of each object.

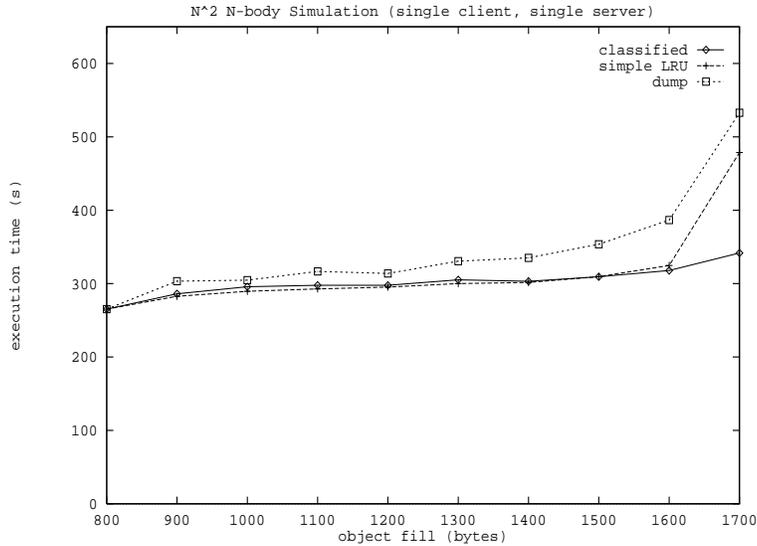

Figure 6.1: Increasing Memory Access Ratio and the $N^2$ $N$-Body Simulation

### 6.1.1 $N^2$ $N$-Body Simulation

Applying the $N^2$ simulation over a range of object fills, for 1500 particles and one simulated time step, yields interesting results. In Figure 6.1 a memory access ratio of 1.0 lies in the object fill range 800 to 900, and a ratio of 2.0 is in the range 1700 to 1800. As expected, higher memory access ratios require more space recovery operations which are reflected in increasing execution times for all three strategies. The application's access pattern generates a constant thrashing of the locally managed memory for memory access ratios greater than 2.0. This occurs because the "current position" particle objects are continually iterated over during the time step simulation, yet a ratio in excess of 2.0 means that this entire set cannot be accommodated in local memory[1].

This experiment shows that identification and preservation of a working set of replicas (classified and simple LRU strategies) yields benefits which increase with the memory access ratio. The observed differences in execution time do not stem from differences in time expended in space recovery operations, these being at most 4 seconds for object fills up to 1600. The time expended in fetching objects is far more significant and varies greatly between strategies. Figure 6.2 shows that, for this application, the number of fetch operations required by the working set strategies is significantly lower than the dump strategy, and that the classified system is better at identifying the working set when the memory access ratio is high. Fewer fetch operations resulting from better identification of the working set translates to significant savings in the time expended processing the fetch operations, as shown in Figure 6.3.

---

[1]The thrashing behaviour actually becomes present at a slightly lower ratio because some objects from the "update" set must also be accommodated.

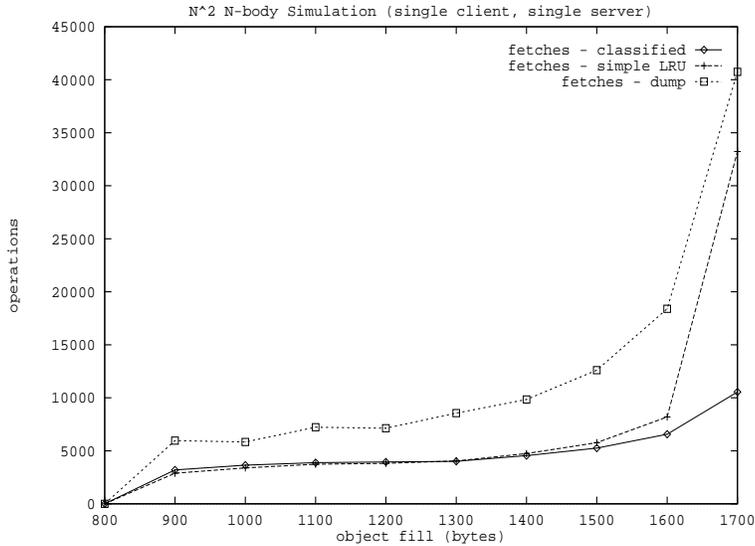

Figure 6.2: Fetch Operations in the $N^2$ $N$-Body Simulation

### 6.1.2 Oct-Tree $N$-Body Simulation

To test whether the observations of the last experiment were consistent across different access patterns, it was repeated using the oct-tree simulation. The oct-tree contains nodes approximating volumes of space in addition to nodes representing the particles. In order to keep the total number of objects comparable between experiments, only 600 particles were simulated, yielding an oct-tree of approximately 3 000 nodes, and this was run through

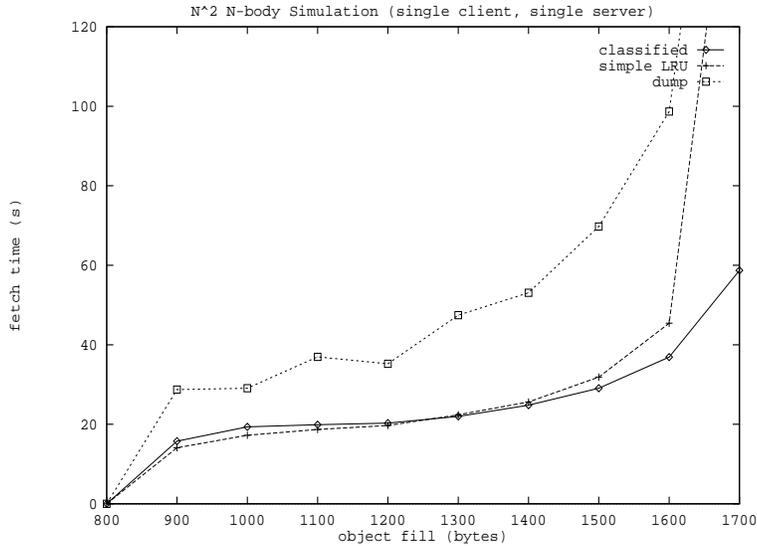

Figure 6.3: Fetch Time of the $N^2$ $N$-Body Simulation

2 simulated time steps.

Figure 6.4 confirms the primary result of the previous experiment, with the increasing memory access ratio (a ratio of 1.0 being achieved in the object fill range 600 to 800, and 2.0 occuring just beyond 1600) showing the dump strategy to be inferior. The latter observations regarding the relative merits of the simple LRU and classified strategies were not confirmed however. For low memory access ratios the classified strategy appears to suit the application's access pattern very well. Increasing the ratio sees the simple LRU begin to give better performance however.

The degradation of the classified strategy's performance results from the classification system interacting badly with the application's data structure and access pattern. To insert a node into the tree the application descends along a path in the tree, inserts a node and modifies nodes as it recurses back up the path. These modified nodes are classified as expensive to discard by the classified system and are kept, at the next space recovery operation, in preference to any clean (and hence cheap to discard) nodes. As the memory access ratio increases the dirty nodes occupy more of the local memory, at the expense of the clean nodes. With each descent path equally likely, the probability of a path requiring discarded clean nodes increases as the ratio increases, hence more fetch requests are generated, and subsequently more space recovery operations (see Figures 6.5 and 6.6). Figure 6.7 supports this hypothesis, indicating that the number of clean replicas discarded by the classified strategy exceeds the number discarded by the simple LRU for larger memory access ratios, this excess being greater than 25% for an object fill of 1600.

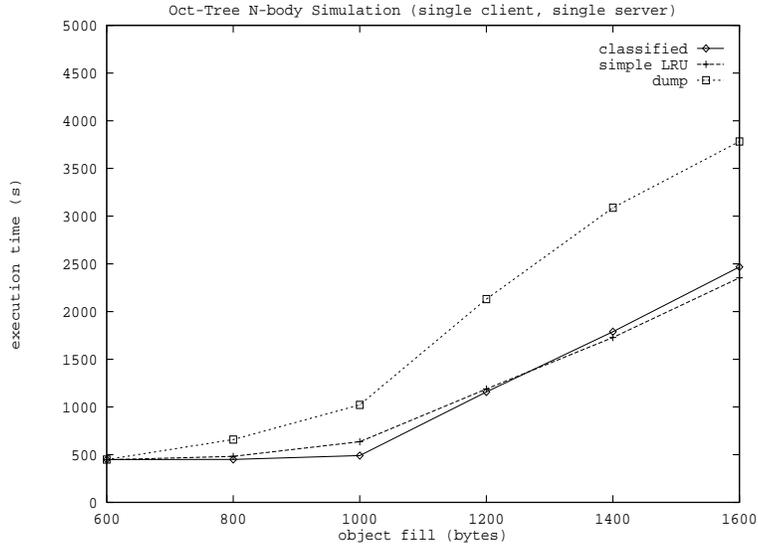

Figure 6.4: Increasing Memory Access Ratio and the Oct-Tree *N*-Body Simulation

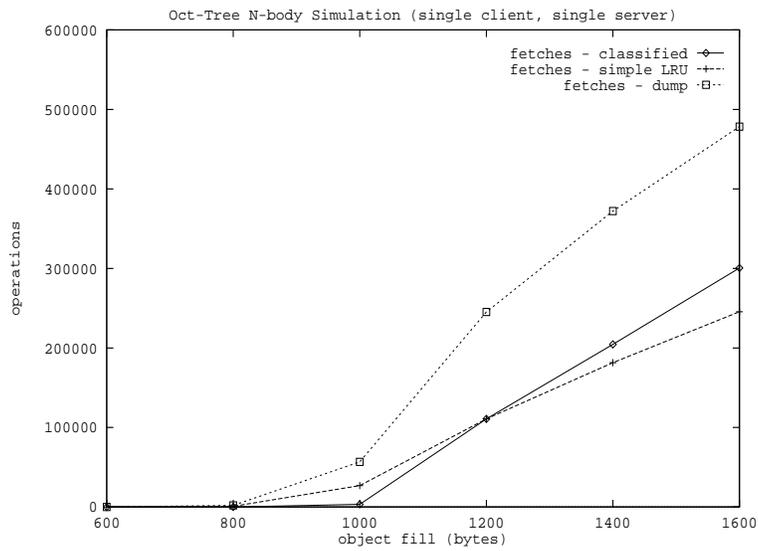

Figure 6.5: Fetch Operations in the Oct-Tree *N*-Body Simulation

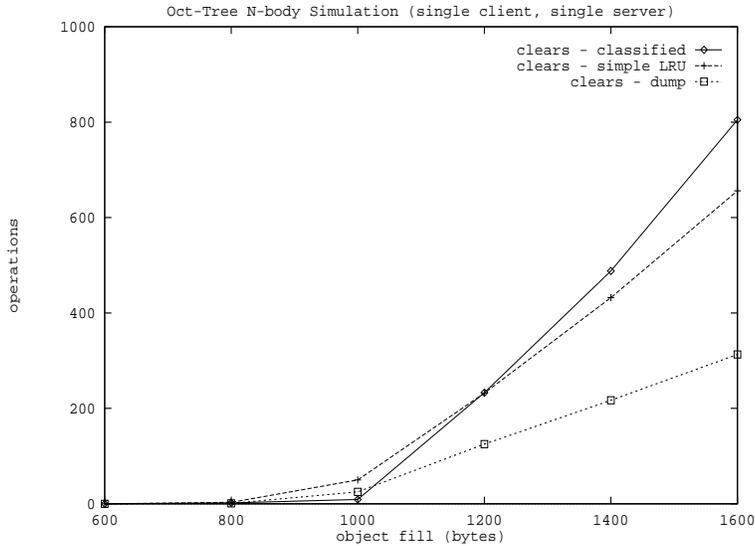

Figure 6.6: Space Recovery Operations in the Oct-Tree $N$-Body Simulation

### 6.1.3   Phased Space Recovery

The object creation phase of the $N^2$ $N$-body simulation involves a series of object creations accompanied by updates of the index objects. The working set is small during this phase, consisting of just the two index objects. It was considered that using the identification and preservation strategies to recover only the targeted amount of memory would be less efficient than removing all replicas and fetching the working set again. Since dump has been shown to be inefficient as an overall strategy, it was hypothesised that the use of different space recovery strategies in different application phases would reduce the cost of individual phases and yield an overall performance gain.

The $N^2$ simulation was run over a range of object fills using the clas-sified space recovery strategy for the time step simulation, and each of the strategies in turn for the object creation phase. The results are summarised in Tables 6.1, 6.2, and 6.3. For any object fill necessitating space recovery operations, the application of the classified strategy to both phases yields performance which is inferior to one of the mixed strategies. In particular, as the memory access ratio increases the use of dump in the object creation phase requires fewer space recovery operations and less total time.

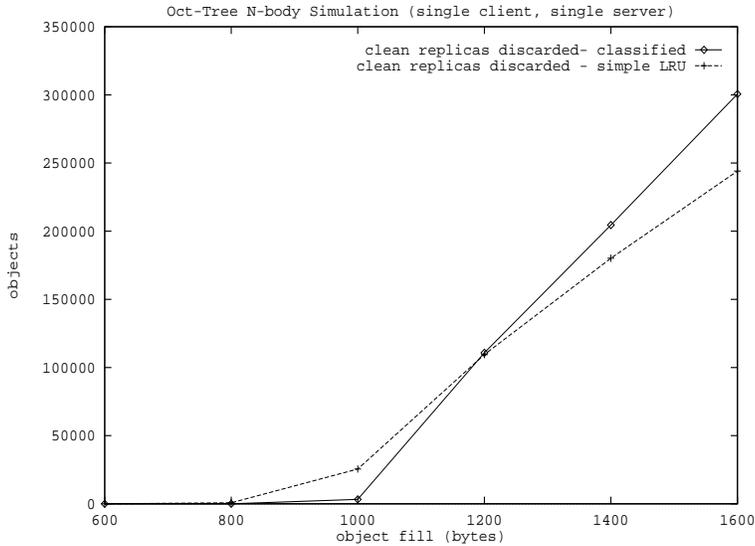

Figure 6.7: Clean Replicas Discarded by the Oct-Tree *N*-Body Simulation

## 6.2  Multiple Servers

To determine the scalability of the space recovery strategies, and the system as a whole, both the $N^2$ and oct-tree simulations were run over a range of servers, with fixed object fills of 1200 and 1000 respectively, and other parameters as detailed above.

It was found (Figure 6.8) that the execution time increases uniformly with the number of servers, for all space recovery strategies. Furthermore, this uniform increase was determined to be due to the increasing cost of creating objects. As noted in Chapter 3, an object creation request which cannot be accommodated by any server responsible for the group will have been passed around all such servers prior to a segment creation order being issued. Thus, the number of messages required to create an object under such circumstances is a linear function of the number of servers responsible for the group. Table 6.4, giving the maximum time expended on any object creation operation, supports this; the data points having a correlation coefficient of 0.9998 and likely to be linearly related.

| object fill | execution time (s) | space recovery operations |
|---|---|---|
| 800 | 265.432 | 0 |
| 1000 | 294.039 | 8 |
| 1200 | 297.545 | 11 |
| 1400 | 301.986 | 15 |
| 1600 | 316.924 | 23 |

Table 6.1: $N^2$ $N$-Body with classified space recovery for both phases

| object fill | execution time (s) | space recovery operations |
|---|---|---|
| 800 | 265.431 | 0 |
| 1000 | 292.3 | 8 |
| 1200 | 296.842 | 11 |
| 1400 | 300.515 | 15 |
| 1600 | 316.33 | 23 |

Table 6.2: $N^2$ $N$-Body with simple LRU followed by classified space recovery

| object fill | execution time (s) | space recovery operations |
|---|---|---|
| 800 | 265.428 | 0 |
| 1000 | 299.131 | 5 |
| 1200 | 297.02 | 7 |
| 1400 | 300.24 | 10 |
| 1600 | 314.694 | 17 |

Table 6.3: $N^2$ $N$-Body with dump followed by classified space recovery

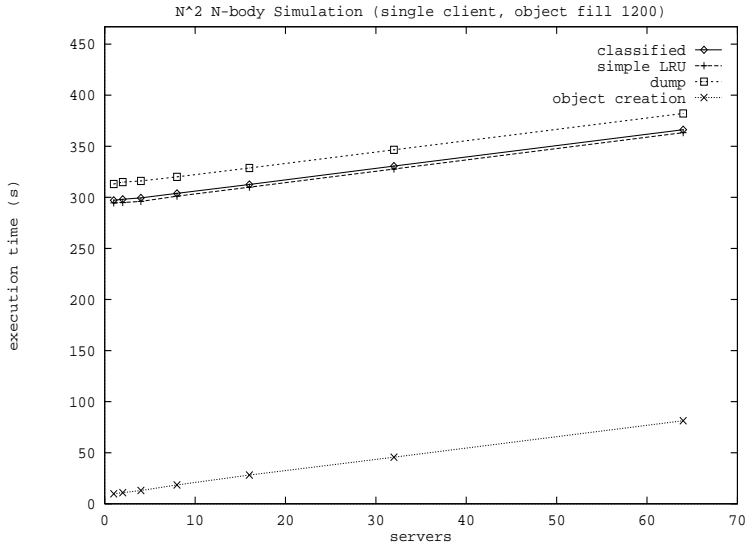

Figure 6.8: Adding Servers to the $N^2$ $N$-Body Simulation

The oct-tree experiment (Figure 6.9) confirmed the general observations noted above. The performance "blip" for the introduction of the second server was found to be present in the total fetch time. Further analysis revealed that the minimum fetch time, and the time expended processing incoming messages (part of the fetch time), exhibit this same blip. When fetching an object, the LOM issues the request and then proceeds to ensure enough free space is available to hold the replica on arrival. If this induces a space recovery operation, servers not occupied with the fetch request may immediately attend any returned replicas and respond with receipts. If these receipts arrive at the client prior to the fetched replica, they must be attended to first and will hence increase the time expended on the fetch.

| servers | max create time (s) |
|---------|---------------------|
| 1       | 0.003186            |
| 2       | 0.003631            |
| 4       | 0.004355            |
| 8       | 0.006056            |
| 16      | 0.009302            |
| 32      | 0.015152            |
| 64      | 0.026994            |

Table 6.4: Maximum Object Creation Times in the $N^2$ $N$-Body Simulation

## 6.3 Multiple Clients

To be more than just a fast single client - multiple server system, it is necessary that the distributed object server be capable of efficiently handling many clients. In order to determine if this was the case with the system implemented, three experiments were conducted. Both $N$-body simulations were tested over a range of client numbers, with the object groups created by each client ranging over all servers (numbering 38 and 43 respectively). Following the observations of these experiments, the $N^2$ $N$-body simulation was rerun using object groups of width one in an attempt to improve performance.

### wide object groups

The initial experiments with widely distributed groups exhibited virtually identical characteristics as the number of clients increased. Figure 6.10 shows the results for the $N^2$ simulation. The execution time increases linearly with the number of clients, with this effect being caused by a serialisation of client requests resulting from bad load balancing on the servers. The synchronisation curve is the time difference between the first client to complete its task, and the last. In a well balanced system, this time would be small for all configurations; in this instance however, it increases with the number of clients. Object creation time, and fetch time (not shown) both increase in a similar manner, suggesting serialisation of requests at the servers. The slow increase in server idle time relative to execution time indicates increased server busy time. Hence, the client execution time increases are primarily due to contention for the servers. It is an artefact of the example application that the clients all work on the same problem, and have their object groups ranging over the same servers. Each client then makes requests on the same servers at the same time. As the servers serialise these requests, some clients will be satisfied prior to others. This ordering is reflected in the time difference for task completion.

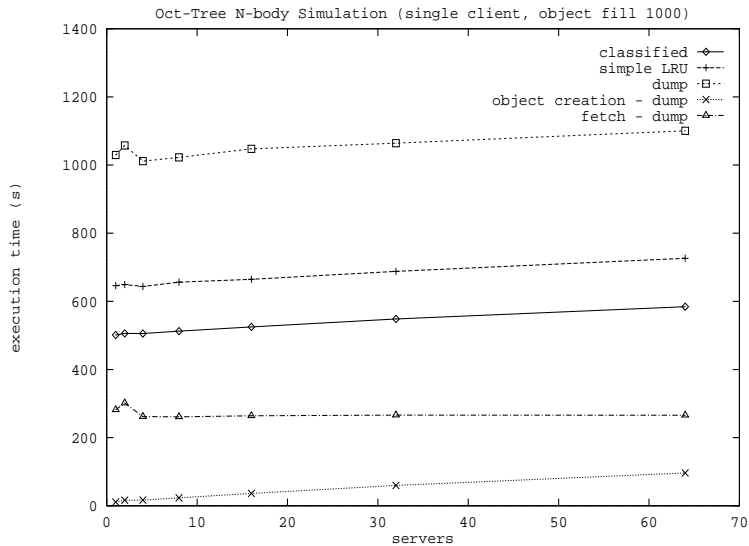

Figure 6.9: Adding Servers to the Oct-Tree $N$-Body Simulation

**narrow object groups**

To overcome this bottleneck, the $N^2$ experiment was performed again, with the object groups having a width of only one server. This resulted in far better load balancing across the servers, with each server responsible for at most $\lceil \frac{\#clients}{\#servers} \rceil$ object groups. The results shown in Figure 6.11 indicate only a small upward trend in the times. This trend is to be expected, because the problems outlined above have not been removed, only minimised. In the previous experiment the serialisation problem was linear with the number of clients, whereas with the narrow object groups of width one, it is linear with the client-sever ratio. More generally, it is linear with the number of object groups for which a server is responsible. It should be noted however, that the example applications may be atypical in having each client working on an identical problem. Applications not having this property would not suffer from the serialisation problem to the same extent.

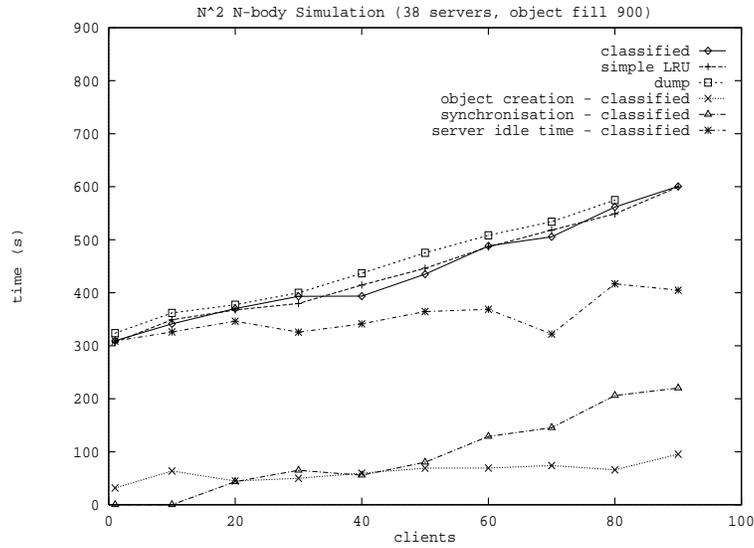

Figure 6.10: Adding Clients with wide object groups to the $N^2$ $N$-Body Simulation

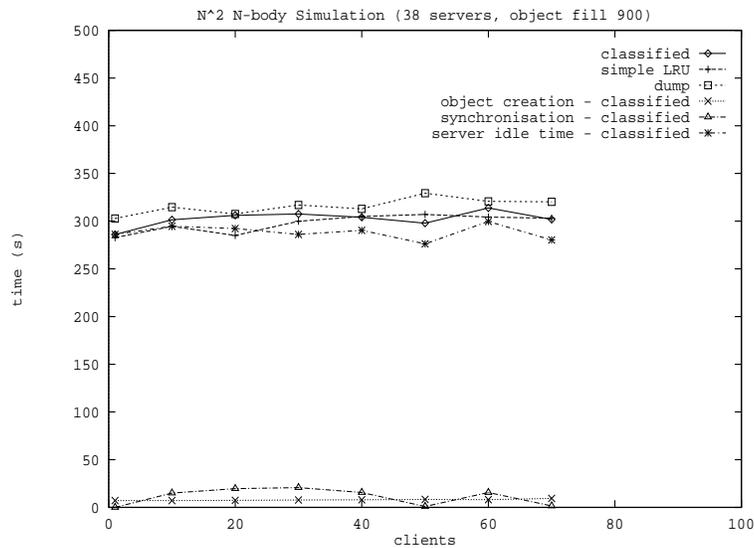

Figure 6.11: Adding Clients with narrow object groups to the $N^2$ $N$-Body Simulation

# Chapter 7

# Object Migration & Coherency

Three mechanisms for migrating object replicas were tested. These were;

- single object upon demand;

- single object upon demand with prefetching to a specified depth; and

- segment (or storage unit) upon demand.

The first of these mechanisms is considered to be the "control", and is the mechanism used in the experiments of the previous chapter. In this chapter, the other mechanisms are evaluated in the context of the binary tree search applications and the oct-tree $N$-body simulation.

## 7.1 Prefetching

The experiments performed to evaluate prefetching as a migration strategy fall into two classes - those aimed at determining the relevant merits of the different methods of prefetching, and those aimed at determining its scalability.

### 7.1.1 benefits

**Binary-Tree Search**

The initial experiment involved the binary tree search application and prefetching to every depth between 0 and 13 inclusive. The object fill was chosen to ensure that space recovery operations, and subsequently object fetching, occured — the fills chosen were 300 and 400 bytes. Both low and high priority prefetch was tested, with low priority tested in both breadth and depth first modes. As can be seen in Figure 7.1, all prefetching methods yielded benefits for both object fills. Prefetching was always beneficial relative to no prefetching for the smaller object fill, whilst this was only true to a depth of 9 for the larger object fill.

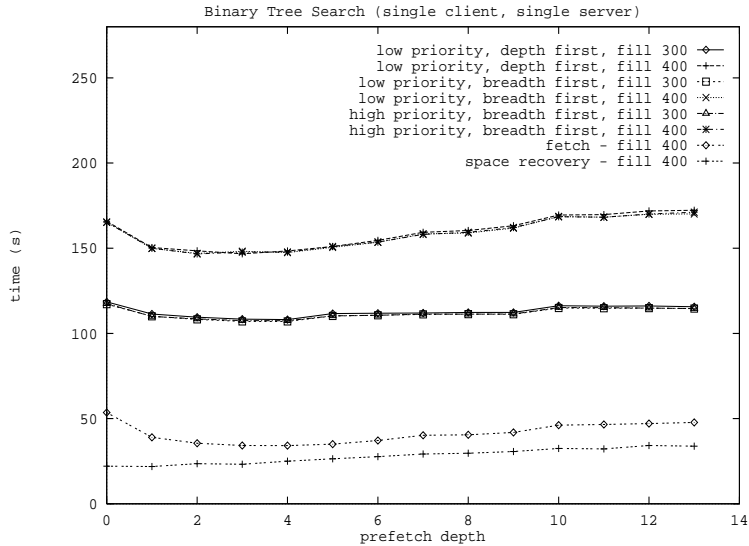

Figure 7.1: Effect of Prefetch Methods on Binary Tree Search Application

That performance declines beyond an optimum depth is due to the changing balance in the tradeoff between the reduction in time expended fetching required objects, and the cost of receiving and processing unused prefetched objects as well as extra space recovery operations induced by their presence in memory. The fetch and space recovery curves in Figure 7.1 illustrate this point. The time expended fetching falls initially and then increases as the processing of an increasing number of prefetched replicas requires more time. It is interesting to note that the number of actual request operations does not increase with the number of space recovery operations. Figure 7.2 shows real request operations to be decaying exponentially with prefetch depth. Unfortunately, an accompanying increase in the average cost of a fetch operation, from 0.004647 seconds at depth 0 to 0.024312 seconds

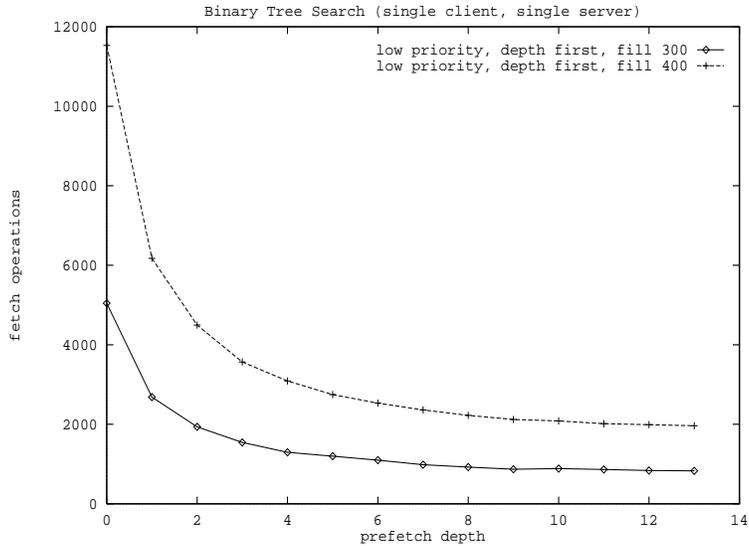

Figure 7.2: Fetch Operations as Prefetch Depth increases

at depth 13 for an object fill of 400, eventually yields increasing total fetch times.

Figure 7.3 indicates that when prefetching larger objects, the number of space recovery operations grows more quickly with prefetch depth. The slow growth in required space recovery operations for the smaller object fill explains the slow rate of performance degradation beyond the optimum depth for the overall execution time on this object fill. That the prefetched objects are responsible for inducing the extra space recovery operations can be seen by noting that the number of prefetched replicas which are accepted into memory and subsequently removed without being used, increases with prefetch depth in a manner similar to that in which the space recovery operations increase (see Figures 7.4 & 7.3).

**Oct-Tree $N$-Body Simulation**

Performing a similar experiment with the oct-tree simulation yielded significantly different results. All prefetch methods were found to give inferior performance for any non-zero depth, as shown in Figure 7.5. Further investigation revealed that the curves for fetch times, fetch operations, space recovery time, space recovery operations, incoming message processing, and

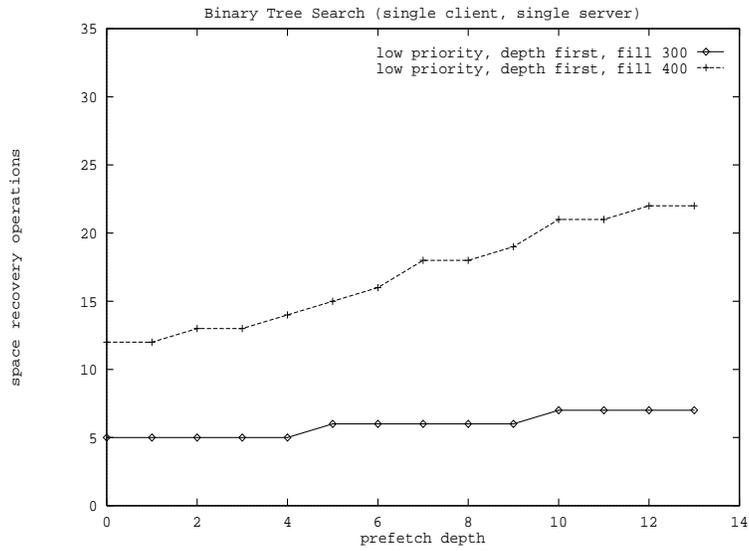

Figure 7.3: Space Recovery Operations as Prefetch Depth increases

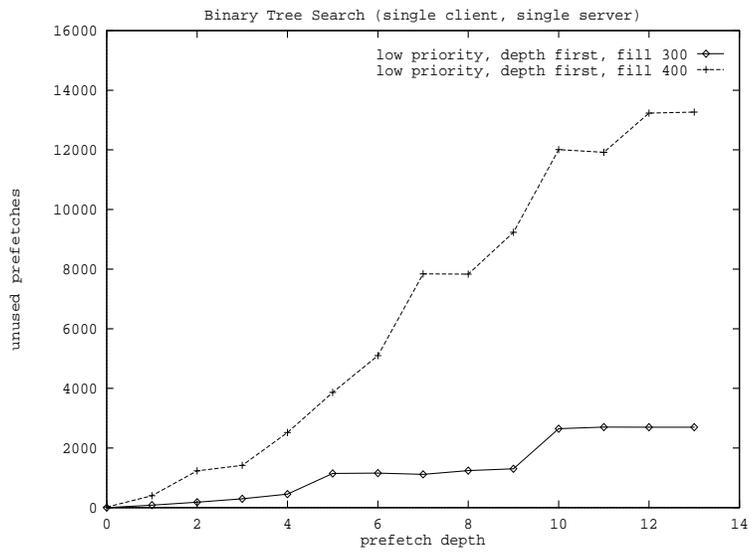

Figure 7.4: Unused Prefetched Replicas as Prefetch Depth increases

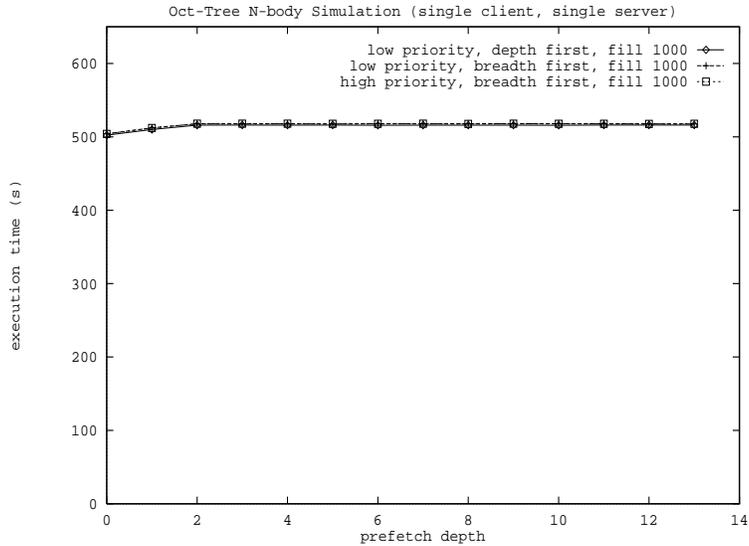

Figure 7.5:  Effect of Prefetch Methods on Oct-Tree $N$-Body Simulation

the number of unused prefetched objects, plotted against prefetch depth, all
possessed the same shape as the execution time curve.  The initial increase in
each curve with the first few depths of prefetch is easily understood in terms
of the prefetched objects (of which there are many when the data structure
has a branching factor of eight) more quickly filling the local cache.  This
necessitates space recovery operations, and generates more fetches because
the working set is not perfectly preserved by the space recovery.  Hence, all
the measurable factors increase initially.

The "flat-lining" of each quantity beyond the initial few prefetch depths
was more difficult to understand.  It arises from the fact that the whole
tree is not accessed when simulating the time steps.  Many of the empty
leaf nodes are not required and hence will never be fetched once initially
discarded.  This allows a complete working set to be constructed in memory.
Once this has been achieved no further fetch or space recovery operations are
initiated.  Additional prefetch depths degrade performance by increasing the
frequency of space recovery operations, or equivalently by bringing forward
the space recovery operations which would be required anyway (and adding
extra operations towards the end of the application as necessary).  Using
the prefetch depth to bring the last operation far enough forward such that it occured prior to the time step simulations, forced the working set to be built early and then remain unchanged. Higher memory access ratios induce more space recovery operations, so it follows that the onset of "flat-lining" should be a function of object fill/memory access ratio. Further experiments exhibited this behaviour and thus supported the hypothesis.

## 7.1.2 scalability

The previous experiments showed prefetching to be beneficial in some circumstances in a single client, single server system. However, the mechanism will only be useful in a distributed object store if it scales well with increasing numbers of servers and clients. Experiments were performed to determine if the binary tree search application continued to show benefits under these variations.

### Multiple Servers

Despite two very different mechanisms being used to implement prefetching (the standard message queue and a separate prefetch queue), both continued to perform well as the number of servers increased. The experiment reported in Figure 7.6 demonstrates that the increasing execution times retain their relative positions as the number of servers increases. With the time expended fetching objects remaining constant, the increase is due solely to the rising object creation time (see §6.2). It was expected that the low priority mechanism would suffer somewhat from the need to send some messages to propagate the prefetch requests between servers. This effect appears to have been either insignificant or counterbalanced by the parallelism of multiple servers attending to requests simultaneously.

### Multiple Clients

To be of value in a truly distributed system, prefetching must continue to yield benefits as the number of client processors increases. Taking the binary tree search with parameters known to yield benefits in the single client, single server system, the application was tested with both wide and narrow object groups over a range of client numbers.

With wide object groups, the server serialisation problem encountered in §6.3 degrades the performance of each prefetch method as the number of clients supported increases (Figure 7.7). As server busy time increases, prefetch requests are either postponed (low priority), or serviced at the expense of real requests (high priority). There are two consequences of this:

- fewer prefetched objects arrive in time to avoid a fetch operation, hence more fetches occur; this necessarily affects low priority prefetching to a greater extent.

- the average time to perform a fetch operation increases; primarily affecting the high priority mechanism.

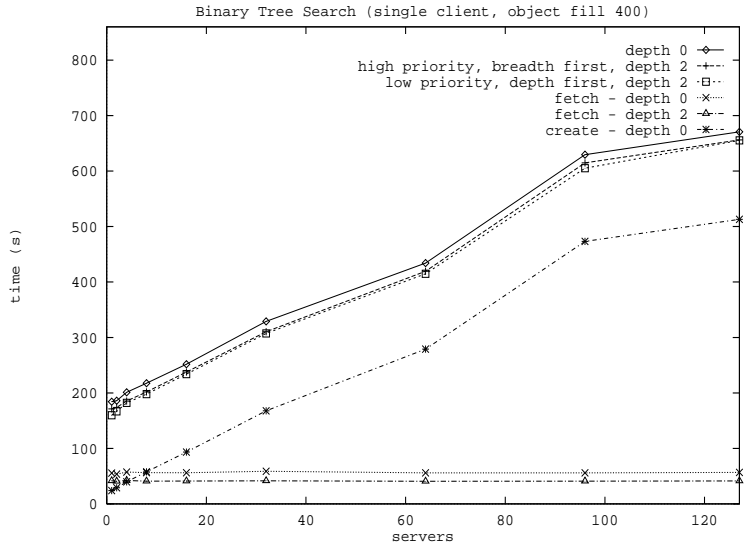

Figure 7.6: Increasing Servers with Prefetch on the Binary-Tree Search Application

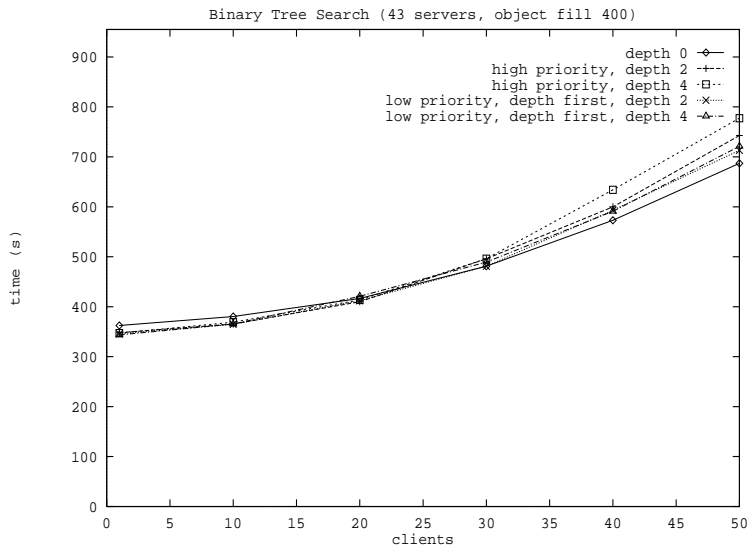

Figure 7.7: Adding Clients with wide object groups and Prefetch to the Binary-Tree Search Application

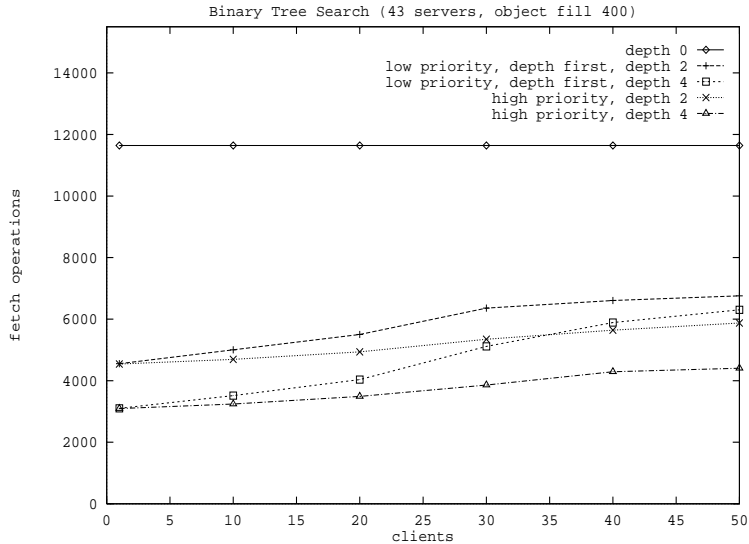

Figure 7.8: Fetch Operations with wide object groups and Prefetch

This view is supported by the curves in Figures 7.8 and 7.9. It is interesting to note that the average fetch time initially falls for low priority prefetching as extra clients are added. Paradoxical as it may seem, this is an artefact of the prefetch requests being postponed for a short time by the servers. A postponement may prevent the prefetched object arriving in time to avoid the fetch operation being commenced. However, it may arrive shortly after and cause the request to be satisfied quickly. Thus, some of the extra fetches indicated for the low priority schemes are quite cheap and can, if not severely outnumbered by extra expensive fetches, lower the average fetch time. This effect is not observed for high priority prefetching, because the prefetch requests are not postponed.

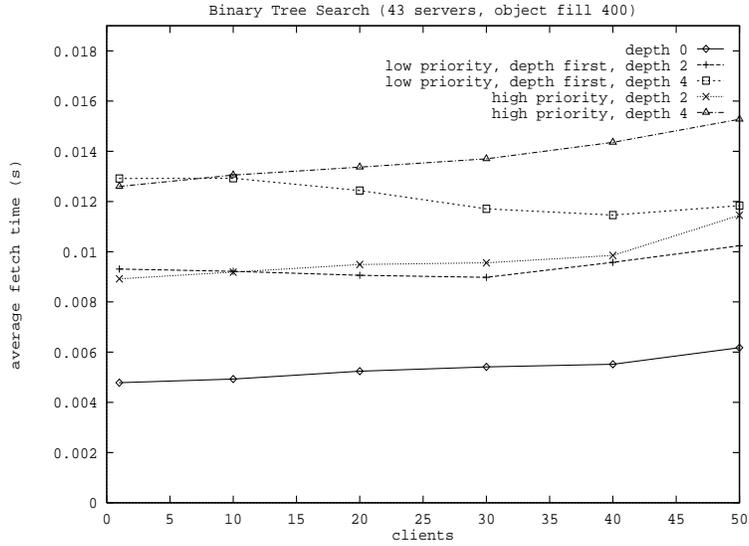

Figure 7.9: Average Fetch Time with wide object groups and Prefetch

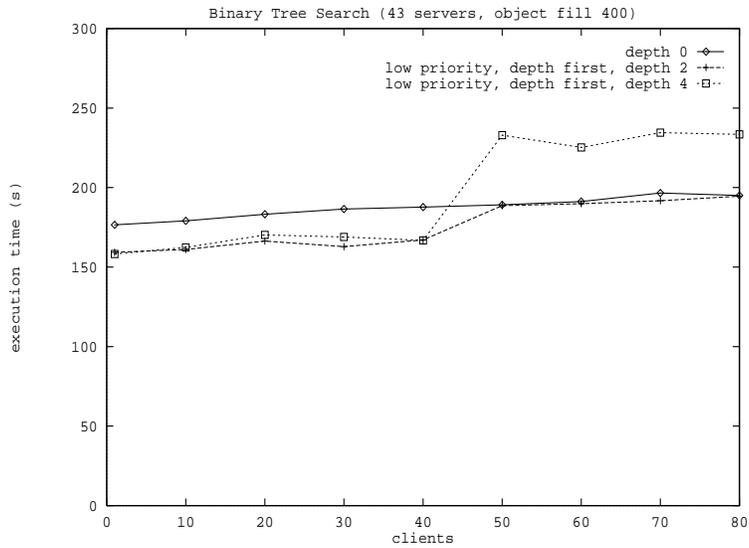

Figure 7.10: Adding Clients with narrow object groups and Prefetch to the Binary-Tree Search Application

As was the case in §6.3, switching to narrow object groups changed the picture significantly when done in the present context. This experiment was performed only for the low priority depth first mechanism. This was because previous experiments had shown no significant difference between mechanisms, the low priority ones having only a slight performance edge if one could be claimed to exist. Figure 7.10 demonstrates that prefetching maintains its benefits until the client-server ratio exceeds 1.0 when a sharp degradation becomes apparent. The degradation occurs abruptly at this point because, for ratios greater than 1.0, at least one server will be solely responsible for servicing the needs of two or more clients. When this happens, the problems outlined above for the wide object group become significant.

## 7.2  Segment Migration

As noted in Chapter 2, migrating objects in their physical storage units has been shown, in the context of disk based object stores, to be far more efficient than migrating them individually. Two experiments were performed to determine if this held true in the context of a multicomputer object server. Within the two $N$-body simulations, the access and storage patterns bear very different relationships to one another. In the $N^2$ simulation, the order of object access within the two sets (current position and update) matches the order in which the objects were created. No such relationship exists for the oct-tree simulation, in which the order of object creation indicates only the order in which particles were inserted into the tree, and does not in any way reflect the manner in which the tree is traversed. In fact, as more nodes are introduced into the tree and the existing leaves are pushed down, objects with the same segment will in fact become further removed in the tree. These examples represent two not unrealistic extremes with regard to the relationship between access and storage patterns, and were chosen to give a broad indication of the relative merits of the two migration methods. Each experiment ran the simulation under both migration schemes over a range of client-server ratios.

In order to minimise the empty space in migrated segments, and hence not disadvantage the segment migration scheme, the object fill and segment sizes were carefully matched for these two experiments. The reasons for this are explained further in Chapter 8.

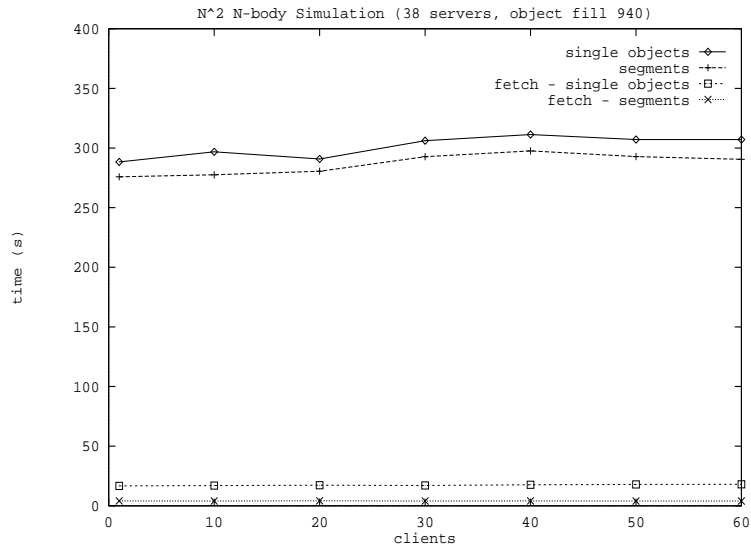

Figure 7.11: Adding Clients to Single Object and Segment Migration: $N^2$ $N$-Body Simulation

### 7.2.1 $N^2$ $N$-Body simulation

For this experiment the problem parameters were, as previously, 1500 particles and a client memory of 3MB. Segments were kept at 50K, but segment directories were reduced to 50 entries, and the object fill selected to completely fill both the directory and the data area of each segment. Narrow object grouping was used to minimise the effect of server serialisation.

It was found (see Figure 7.11) that migration by segment possessed a consistent advantage of approximately 4%, over single object migration throughout the range of clients. This resulted primarily from a 98% reduction in the number of fetches which yielded a 75% reduction in the total fetch time. The larger granularity of segments was reflected in a larger average fetch time, but also in a reduction of the number and cost of incoming messages. Table 7.1 gives comparative figures for the single client case.

### 7.2.2 Oct-Tree $N$-Body simulation

Attempting to conduct this experiment with the previously used oct-tree problem parameters (3MB client memory, 600 particles, 2 time steps, and object fill 1000), served to demonstrate the great differences between the two migration schemes. Across the client range the object migration scheme completed execution in approximately 8 minutes, whilst the single client case of the segment migration scheme failed to complete within 30 minutes when the process was terminated. In order to gain comparative figures, and provide a lower bound on the relativities between the two schemes in this instance, two reductions were made to the size of the computation required. The time steps were reduced from 2 to 1, and the number of particles to which the simulation was applied was reduced to 10 from the full 600 (note that the oct-trees were still constructed using all 600 particles). As with the last experiment, the number of directory entries in each segment was reduced, on this occasion to 40, and the segment data area was reduced to 49 440 bytes.

| migration scheme | object | segment |
|---|---|---|
| execution time (s) | 288.34 | 275.88 |
| fetches | 3 507 | 73 |
| mean fetch time (s) | 0.00475 | 0.05666 |
| total fetch time (s) | 16.67 | 4.13 |
| space recovery ops | 7 | 6 |
| mean space recovery time (s) | 0.72 | 0.73 |
| ROT searches | 45 523 | 18 052 |
| total ROT search time (s) | 1.70 | 0.74 |
| incoming messages | 3 524 | 126 |
| total incoming message time (s) | 4.27 | 2.01 |

Table 7.1: Single Object vs Segment Migration for the $N^2$ $N$-Body Simulation

Even for this reduced simulation, Figure 7.12 shows the segment migration scheme to deliver vastly inferior performance. Increased fetch time accounts for some of the difference, segment fetch time actually exceeding single object execution time. Neither scheme was able to successfully complete the simulation for a client-server ratio in excess of 1.0 due to memory limitations on the server processes[1]. From Table 7.2 it can be seen that the

---

[1]Each cell has 16MB of memory.  In this instance the server processors utilise it as follows:

- operating system 4MB
- distributed server executable 1.2MB

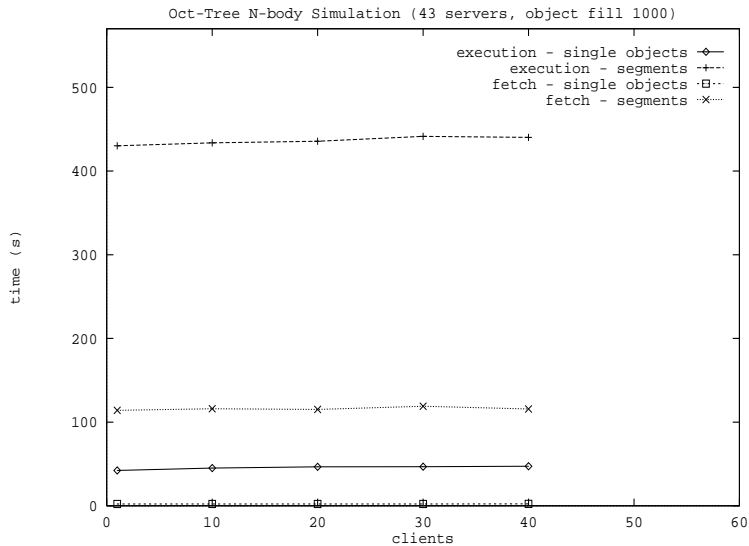

Figure 7.12: Adding Clients to Single Object and Segment Migration: Reduced Oct-Tree *N*-Body Simulation

segment migration scheme requires five times the fetches of the single object scheme, 42 times as many space recovery operations, and as a consequence of these, a factor of 48 more searches of the ROT. These increases result from a thrashing of the client memory which occurs because the segment migration scheme actually reduces the effective amount of client cache. With segment migration, accessing an object requires the segment containing the object to have a replica in the client cache. As noted above, objects which are closely related in the oct-tree are unlikely to be stored in the same segment. This effectively increases the size of each object in the client cache to the size of a segment, hence fewer useful objects can be stored in the cache. Additionally, fetch operations are not only more frequent, but also more expensive because larger amounts of data have to be moved.

- segments and supporting data structures (for 2 clients) 10.7MB

This leaves little space for message buffers, with just those required to service a single client undertaking a space recovery operation likely to cause the server to exhaust memory.

| migration scheme | object | segment |
|---|---|---|
| execution time (s) | 42.22 | 430.16 |
| fetches | 459 | 2 210 |
| mean fetch time (s) | 0.00481 | 0.05159 |
| total fetch time (s) | 2.21 | 114.02 |
| space recovery ops | 4 | 169 |
| mean space recovery time (s) | 1.16 | 1.61 |
| ROT searches | 7 938 | 381 539 |
| total ROT search time (s) | 0.28 | 14.72 |
| incoming messages | 473 | 2 431 |
| total incoming message time (s) | 0.62 | 50.19 |

Table 7.2: Single Object vs Segment Migration for the reduced Oct-Tree $N$-Body Simulation

## 7.3 Concurrency

The coherence maintenance scheme implemented was object locking based upon binary semaphores. Experiments were conducted to determine the cost of this scheme for two applications with different usage patterns. The PLUCK string simulation exhibits a pattern in which few objects are concurrently accessed, and this is done in a predictable manner. In the parallel oct-tree application, all objects are potentially accessed concurrently necessitating greater coherence maintenance expense. Both applications are fully described in Chapter 5, and the experiments performed are described below.

### 7.3.1 PLUCK String Simulation

For this experiment the PLUCK application was run for up to 90 clients supported by 37 servers. Fifty thousand points were simulated for twenty time steps, and the client replica cache memory was increased to 10.5MB to avoid the complication of space recovery operations. Since the setup phase involves no significant concurrent access, it was excluded from calculations and only the actual simulation phase considered. All replicas required by a client were warm in the local replica cache at the commencement of simulation.

It was found that the simulation time followed the shape $k + \frac{1}{x}$, where $k$ is a constant and $x$ is the number of clients (see Figure 7.13). This curve shape is not unexpected, similar experiments with message passing versions of this application have shown identical behaviour. Figure 7.13 also shows the coherence maintenance cost (time expended on Waits and Signals) to have the general shape $k(1 - \frac{1}{x})$. In the single client case, there exists no competition for the semaphores, or for the servers' attention. Introducing multiple clients introduces contention for semaphores and server processing. Hence the steep initial rise in the coherence maintenance cost, and subsequently slower growth as extra clients are absorbed by different servers, and add less to the higher base. The greater the number of clients, the greater the number of objects being concurrently accessed, and the greater the contention. The upper bound for this curve occurs when all points being simulated are concurrently accessed. It is believed that the slight falls in the coherence maintenance cost for higher client numbers are due to cache effect — for such large client numbers the number of object replicas per client is small, and is likely to, along with supporting data structures, fit within the 128KB cache of the AP1000 cells. The message passing version of the application maintained constant coherence maintenance cost (communication cost) since processors communicated directly rather than through intermediary servers.

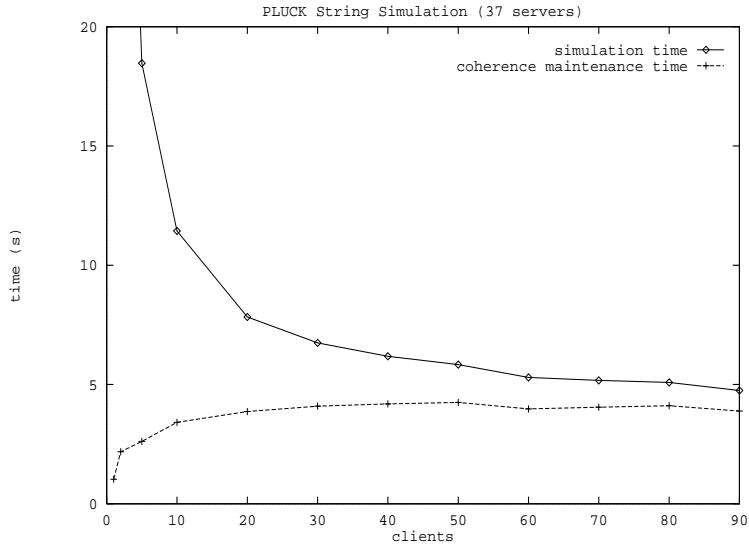

Figure 7.13: Adding Clients to the PLUCK String Simulation

Figure 7.14 compares the efficiency of the distributed server version of the application with message passing versions. It is worth noting that whilst the distributed server version possesses the lowest efficiency curve, the curve is still of the general shape exhibited by the message passing version. The efficiency of the message passing program simulating 50 000 points probably represents a practical upper bound not attainable by an application supported by a general object management system. It must be noted that the message passing program used the CELLOS message passing routines, and not the slower PVM library with which the distributed object server was implemented. Changing the message passing library for these applications would undoubtedly result in different efficiency curves, but should not alter the general shape of the curves.

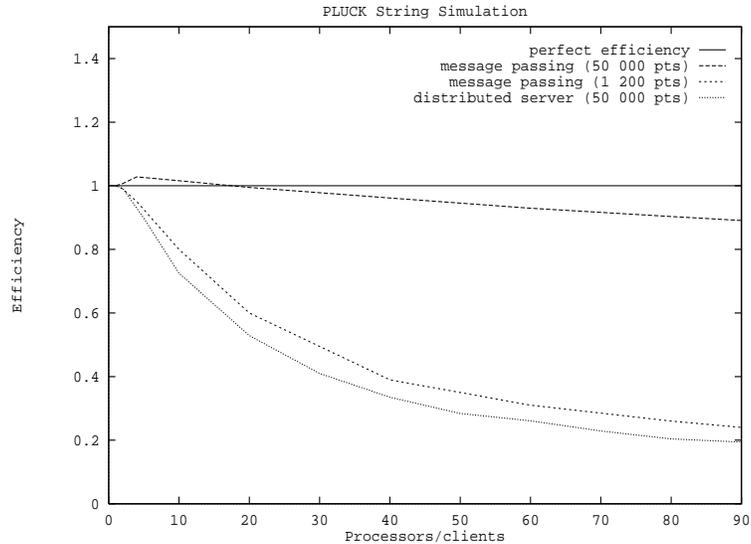

Figure 7.14: Efficiency of PLUCK String Simulation

### 7.3.2 Parallel Oct-Tree $N$-Body Simulation

This experiment saw the parallel version of the oct-tree $N$-body simulation run against up to 90 clients, supported by 37 servers. Six-hundred particles were simulated through a single time step, and as with the last experiment, a client replica cache of 10MB was used. All clients had an empty replica cache at the commencement of the simulation phase.

In Figure 7.15 it can be seen that for small numbers of clients the time expended on Waits and Signals is a significant proportion of the overall execution time. If the non-parallelised setup phase is excluded however, the cost of maintaining coherence is found to increase linearly as a proportion of the parallelisable execution time[2] — from 81.66% for a single client to 92.31% for 90 clients with correlation coefficient 0.972. Whilst the number of object accesses, and hence the number of Waits and Signals falls as $\frac{1}{x}$ with the number of clients, the average Wait time increases linearly as contention both at the server and for object semaphores increases (see Figure 7.16; data points have a correlation coefficient of 0.979). This does not compare well with message passing versions of the same application which achieve coherence maintenance costs of approximately 15% for 2 560 particles over 128 processors. With such large proportional coherence maintenance costs for the object server version, the 11 percentage point increase over the given range has dramatic effects on efficiency, as shown in Figure 7.17.

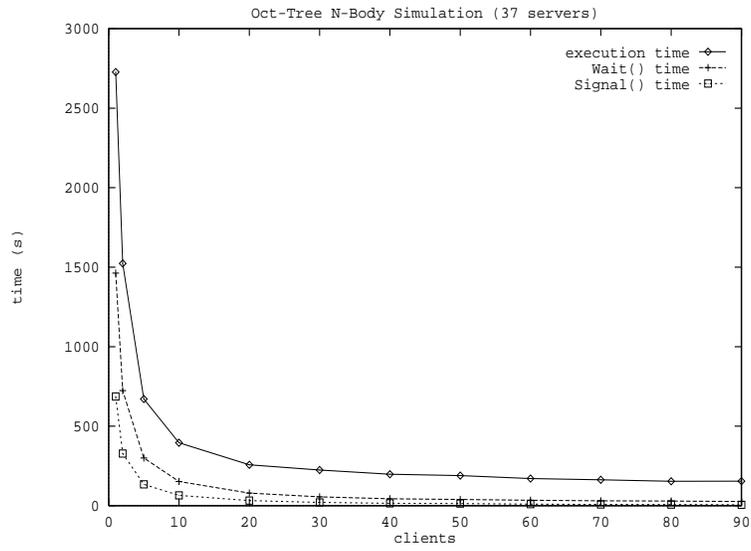

Figure 7.15: Parallel Oct-Tree *N*-Body Simulation

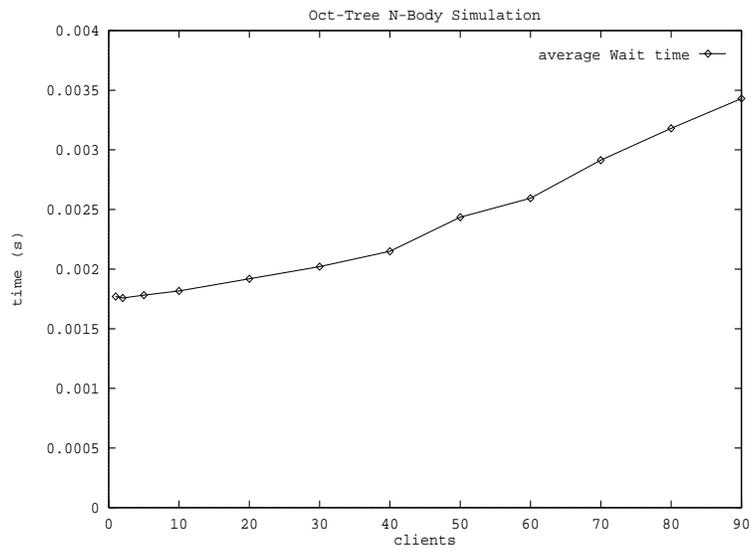

Figure 7.16: Average Wait Times of Oct-Tree *N*-Body Simulation

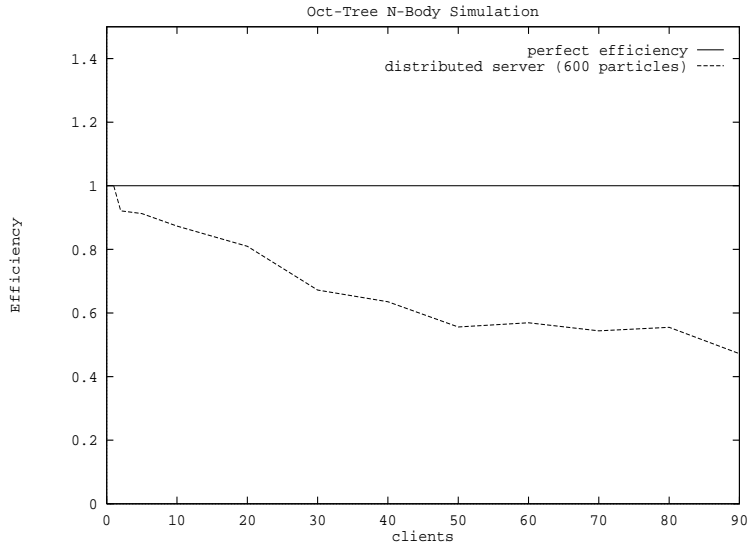

Figure 7.17: Efficiency of Oct-Tree *N*-Body Simulation

---

[2]This is the simulation time less the fetch time. Each client starts the simulation with a cold replica cache, hence adding clients increases the total amount of fetch work performed.

# Chapter 8

# Discussion

This chapter discusses the experimental results of the previous two chapters in the context of the issues falling within the project scope. The general performance of the system is discussed, including the merits of the ROT and the various space recovery strategies. This is followed by a comparative evaluation of the object migration schemes investigated, and a discussion of the performance of the concurrency control mechanism. Finally, there is a discussion of the implementation platform in relation to the system's performance.

## 8.1   General Performance

With the exception of object creation cost, the experiments presented in Chapter 6 demonstrated performance of the distributed object server to scale in an acceptable manner with respect to the tested criteria — memory access ratio, number of supporting servers, and number of clients.

### Object Creation

As previously noted, the object creation mechanism in the GOM is responsible for the object creation time scaling linearly with group width. Since the present system does not allow deletion, the searching of all relevant servers for free space prior to the creation of a new segment could be omitted. In a system permitting object deletion a means of locating and reusing this space is however, necessary. One solution is the maintenance of a free space list by the base server of each group, detailing the largest amount of free space within a segment possessed by each of the other servers responsible for the group. This would allow immediate detection of the need for a new segment to be created, and otherwise allow the creation request to be directly forwarded to a server capable of completing it.

## Resident Object Table

The ROT structure performed as expected, giving access times increasing with the log of the number of entries, the maximum access time for a table of 3000 entries being 0.0001 seconds. Whilst this is time efficient, the space efficiency of the structure is questionable. Effectively maintaining two structures (the RED-BLACK tree, and MRU list) necessitates a minimum of five pointers per entry. To improve performance seven are actually used. For a ROT containing 4000 entries, pointers alone occupy 112KB, which represents 4% of the replica cache with which the sample applications were run. For larger memory access ratios, which often result from a large number of objects, a significant improvement in performance may be gained if this memory were to be used to cache replicas. The linked list is already known to be inadequate, but a large hash-table using chaining to resolve collisions, as used in Mneme[27] and PS-Algol[9, 4], may have access times not significantly different from those of the RED-BLACK tree. A balanced binary tree containing 10 000 nodes has a greatest depth of 14, whilst a table with 1 000 hash values can contain 10 000 elements with an average chain length of 10. Such a scheme would have a higher fixed overhead, but a lower marginal overhead, and represents a trade off of access time scalability for space scalability.

## Space Recovery Strategies

The experiments uniformly showed the dump strategy to be the single worst strategy for complete applications, with no clear distinction between the simple LRU and classified strategies. This last result is the more interesting, since at face value the two schemes are very different. The primary ordering of the simple LRU when discarding objects is just that obtained from the LRU threading of the ROT, whilst the classified scheme is ordered primarily upon the four classes (nominally eight, but reducing to four if segment migration is ignored) and by LRU within these. It appears this different primary ordering is often not significant. In fact, some argument can be made for the two total orderings being similar. Objects for which there are no swizzled local references, by definition, have not been accessed since the
last space recovery operation. Hence, they should appear near the LRU end
of the MRU→LRU threading and be discarded early in the process by the
simple LRU strategy. It is however, precisely these objects which compose
the first two classes of objects (or segments) discarded under the classified
strategy.

As can be seen in Figures 6.1 and 6.4, the classification scheme deviates
from the simple LRU at higher memory access ratios. This occurs because
the primary ordering of objects by class, in the sequence implemented, differs
from the LRU information in a manner which either more closely matches
the true working set ($N^2$ $N$-body simulation), or less closely matches it
(Oct-Tree $N$-body). This suggests that classified strategies with different
class orderings may be better suited to individual applications, and further
supports the result of the phased space recovery experiment (§6.1.3). That
no one space recovery strategy is best in all cases should not be surprising.
Each attempts to predict which replicas to discard and retain so as to min-
imise the total number and cost of space recovery operations, and hence the
applications execution time. To achieve optimum results in all cases requires
correctly predicting the application's subsequent access pattern, which can
only be done by simulating the application's continued execution. This is
not a viable option, so space recovery strategies are based upon certain as-
sumptions, the application's previous object accesses, and the replicas in
local memory. A strategy's assumptions may or may not hold for a par-
ticular application, or even for a particular phase of an application, hence
different strategies will exhibit varying relative performances on different
applications.

## Client-Server ratio

Whilst Figures 6.10 and 6.11 show server idle time to be significant pro-
portions of total execution time, they also demonstrate the impact of bad
load balancing across the servers. With wide object groups, and the clients
all running identical problems, all clients make the same requests of the
same server at once. Serialising these requests, the server imposes an order-
ing upon the clients which may continue through subsequent computations.
Improving the load balance by using narrow object groups, and hence hav-
ing each server solely responsible for only a few clients, is a solution which
scales far better in this case. This may however, not be true in the more
general case where each client does not work on an identical problem. If different clients exhibit different usage patterns for the objects with which they work, defining narrow object groups runs the risk that the groups for which a particular server is responsible will be more heavily used than those of another server. This includes the possibility of a narrow group being heavily used by many clients, and would lead to a bad load balancing which could be remedied through the use of wide groups.

In the present system an object group is limited in size by the memory available on the servers responsible for it. This is linearly related to the width of the group, hence wider groups have a higher bound on their size. This restriction would disappear if persistence were added to the system, and the memory in the servers seen as a cache for the stable store. However, wider groups would then benefit from larger distributed caches and the interaction between group width and client-server ratio would still be relevant.

## 8.2 Object Migration Schemes

Object migration is expensive, the aim of every migration scheme is to minimise the this cost. The default strategy in the implemented system is single object migration on demand. This migrates only those objects which clients immediately require in their working sets. The two other schemes evaluated in relation to this default were prefetching and segment migration. Each of these attempts to reduce the cost of object migration to the client by predicting which objects the client will subsequently request in order to complete its working set. The prefetch mechanism assumes clients will construct a working set consisting of the transitive closure of the object it fetches, whilst segment migration works on the assumption that storage locality matches temporal access locality.

### Prefetching

The experiments of §7.1 serve to illustrate the large number of factors affecting the degree of any benefit gained from prefetching. It is clear that for each application tested, there existed an optimum prefetch depth (possibly zero) dependent upon the data structures branching factor, the memory access ratio, and the client-server ratio. Increasing any of these factors or the prefetch depth degrades performance. Increases in the client-server ratio, with wide or narrow object groups introduce contention for the processing of requests at the servers which can be minimised by the use of appropriate object group widths, but not eliminated. Extra client operations are induced by increases in the prefetch depth, branching factor, and memory access ratio. As the client replica cache is more quickly filled by prefetched objects, space recovery operations, and subsequently extra fetch operations, are induced. Whilst the prefetching does construct the working set with fewer fetches, the cost of these induced operations soon outweighs the savings. Better performance could be gained from prefetching by reducing the effective branching factor and/or prefetch depth through more accurate prediction of objects which will form part of the client's working set.

### Segment Migration

As mentioned in §7.2, the object fill and segment sizes were carefully matched for the segment migration experiments. This was done so as to not disadvantage the segment migration scheme by migrating segments containing empty space. The present implementation has a fixed directory size and fixed data area for each segment. When one of these becomes full, the space remaining in the other is wasted, and reduces the effective replica cache size when segments are maintained within it. A more space efficient scheme is known, in which the directory and data areas are of variable size, and grow from opposite ends of a piece of memory. This scheme wastes little space, and would therefore be advantageous to the segment migration scheme. The matching of the object fill and segment size, such that the size of a segment directory is exactly the number of objects which can be accommodated in the segment data area, gives no space wastage and hence simulates the known better scheme.

The segment migration experiments of §7.2 yielded results which appear contradictory to those of Hosking and Moss[15] who report speedups of up to 72 for fetching, and 76 for data structure traversal when migrating storage units rather than single objects. The $N^2$ $N$-body simulation experiment revealed a speedup of only 1.04 for segment migration over single object migration, and the oct-tree experiment showed segment migration to actually be much slower than migration by single object. There are several reasons for these discrepancies. Primarily, the Mneme persistent object store[27] used by Hocking and Moss does not possess the equivalent of dynamic space recovery operations [1] , hence there is little cost in migrating segments containing objects which the application does not use. This highlights a major architectural difference between the AP1000 and the workstation cluster on which Mneme runs. AP1000 processors have only 16MB of physical memory and no virtual memory, and thus must perform space recovery operations if they are to work with sets of objects larger than the local replica cache. Space recovery was a significant portion of the difference between the two schemes in the oct-tree experiment, and generated many extra fetches for the segment migration scheme. Additionally, the sample applications on which Hosking and Moss base their times are but simple benchmarking programs which fetch a number of random objects or transitive closures and perform null procedures on each object. Fetch operations easily dominate the profile of these programs; combining this with the lack of space recovery operations, their timings exaggerate the benefits of segment migration. Consider the $N^2$ experiment which favoured segment migration — the total time expended fetching indicated a speedup of 4.02 with segment migration. In the context of the application as a whole, this was reduced to just 1.04. In the reduced oct-tree application where each segment fetched contained mostly unwanted objects, the cost of space recovery operations and induced extra fetches gave appalling segment migration performance. Hosking and Moss do not allocate newly created objects to segments until the objects are first removed from the client's address space. This is done in a breadth first manner. Given the presence of space recovery operations, and the manner in which the oct-tree is constructed in the $N$-body simulation, it is believed that such a method would not yield a qualitatively different distribution of objects across the segments for the oct-tree application.

---

[1] "We simply allocate virtual memory at will and write modified segments back only when their file is closed." [27]

It should be noted that both the results of Hosking and Moss[15], and those presented here, were obtained by running applications using many small objects per segment. Hosking and Moss used segments containing up to 480 objects of 68 bytes. A minimum of 40 objects per segment was used in the experiments present in the previous chapter. Whether these results hold for larger objects and/or multi-segment objects is a question yet to be addressed.

An additional significant difference between the environment in which the Hosking and Moss' results were obtained and that used for those presented here, is the existence of a remote backing store in the Mneme system. This was an extra, and necessarily slow layer, of data migration within the system. Only storage units were migrated between the stable store and the Mneme equivalent of the combined storage layer and global object manager. In light of the high latency and low throughput of an Ethernet channel to a remote database server, this is a reasonable strategy. Such a strategy would probably be useful for the addition of a stable store to the distributed ob-ject server presented here. As noted previously, it would be hoped that the system would then benefit from the large cache memory which the servers would become.

## 8.3 Concurrency

The binary semaphore on each object implemented as a coherence control measure is essentially the same as the *volatile pools* approach proposed for, but not implemented in, Mneme[27][2]. The claim of Moss that this mechanism is sufficient to support the construction of any desired concurrency control semantics is undoubtedly true. On the basis of the experiments in §7.3 however, the suitability of this mechanism for concurrent access to distributed data structures must be questioned. In both experiments the coherence maintenance costs quickly became the majority of the simulation time and subsequently impacted heavily on efficiency. For just a single client running the oct-tree simulation, the assumption that coherence maintenance using semaphores was necessary cost approximately 2 200 seconds. On the efficiency scale used in Figure 7.17, the non-parallel version of the oct-tree application (that is, a single client version which assumes no concurrent access and uses no semaphores), rates as 550% efficient! Given that the simulation of a time step in the oct-tree application can be broken into distinct read-only and update phases, it is inefficient for the coherence maintenance mechanism to support only exclusive access to objects. A crude form of multiple-reader access was constructed using the dump and synchronise operations. This improved performance significantly, reducing the single client execution time to under 600 seconds. This is however not a general solution, and should be considered only as an example of the degree of performance which it is possible to attain.

A second problem with the binary semaphore mechanism which became apparent during construction of the example applications, was the difficulty of programming with it. Each access to an object which may be concurrently accessed must be enclosed within a Wait and Signal on that object. Importantly, the Wait and Signal operations must be nested in such a manner as to maintain consistency across any objects being updated, and avoid deadlock. Whilst not difficult for small examples like the PLUCK simu-lation, it is still a non-trivial task. In contrast, personal experience with object-oriented databases has shown the transaction model to be intuitive and readily applicable to substantial applications. However, the provision of synchronisation between client processors, as required in PLUCK, using transactions alone is difficult.

---

[2]This equivalence assumes that the application programmer is correctly using the binary semaphores to ensure mutual exclusion. In the absence of this, the system makes no guarantee of coherence maintenance under concurrent object access.

## 8.4   Implementation Platform

The implementation platform, both hardware and software, influenced system performance in several ways. Most notable was the slowness of the PVM message passing library, and limited processor memory.

Whilst PVM facilitated development greatly by allowing most software to be tested on a network of workstations prior to use on the AP1000, it is known to be relatively slow compared to the machine's native CELLOS message passing routines[19]. A ping-pong benchmark[3] of PVM yields a time of $1.35 \times 10^{-3}$ seconds. This represents 71% of average object creation time under optimal conditions[4]. Increasing the ping-pong reply message to 1000 bytes yields a time of $1.80 \times 10^{-3}$ seconds, representing 42% of average fetch time for objects of this size. This indicates that communication is a substantial overhead on these operations. With create and fetch time together contributing the bulk of the system time in most of the cases tested, it is clear that a reduction in the point-to-point communication time would result in greater system performance. The low latency line-sending option available on the AP1000 was investigated and found to be severely restricted by the 512KB limit on the ring buffer in which incoming messages are stored. This limit was regularly exceeded with just a few client processors returning replicas to a server after space recovery operations, or by clients working with object migration strategies such as prefetching or segment migration. Line-sending is hence, not an optimisation that can be generally applied in this context. A communication optimisation likely to be generally applicable and yield better performance than line-sending alone would have, is the replacement of PVM as the message passing library with either CELLOS or MPI[12]. All communication in the distributed object server is achieved by macros which call the appropriate PVM routines, because of this it is believed that a change in the underlying message passing library would not be difficult.

---

[3]This is the time taken for a zero length message to be sent from one processor, be received, and a reply sent and received.

[4]Single client, single server, and no immediately preceeding space recovery operations.

As noted in §7.2.2, the 16MB per processor physical memory limit was a problem for higher client-server ratios. Together the operating system and distributed object server code occupy about one third of the available memory. With a large amount of space required for storing segments from the shared distributed heap and/or large communication buffer requirements, 16MB can be easily exhausted. Both contributing factors are exacerbated by higher client-server ratios and narrow groups. A possible solution is the provision of stable storage for the object server and allowing servers to store infrequently used segments on this medium. The object server would then become a large distributed cache for the stable store.

# Chapter 9

# Conclusions

How best to use distributed memory multicomputers to enhance the performance of data hungry applications is an open question. Utilising the large distributed memories of such machines as a distributed cache for a stable object store is a feasible solution. The issues involved have been explored via the construction of a distributed object server on the AP1000. This system has been used to investigate solutions to the problems of space recovery, object migration, and concurrency management.

## 9.1 Conclusions

The implemented distributed object server supports extensions to the C++ language allowing objects to be created and manipulated in a shared distributed heap. The processors of the AP1000 are divided into two groups, clients and servers. Server processors store and manage the distributed heap, while client processors run application programs and access the heap through library calls. The construction and subsequent instrumentation of this system whilst running several example applications has given valuable insights into several of the issues concomitant with a multicomputer object store.

### 9.1.1 General Performance

Under different example applications, it was observed that the distributed object server scaled in an understandable manner for increases in the memory access ratio, the number of servers, and the number of clients. Of the space recovery strategies tested, dump was found to be the worst performer in the general case, with no clear distinction between the classified and simple LRU strategies. Each strategy was determined to be optimal under some circumstances. The phased space recovery experiment (§6.1.3) demonstrated that dynamic control of the space recovery strategy by the application program can produce performance benefits.

In systems which permit object deletion or garbage collection, the creation of a new object is potentially expensive. The implemented method for detecting free space scales linearly with the width of an object group and adds considerably to the expense of object creation in wide object groups. This behaviour strongly warrants the investigation of either different free space detection techniques or other object creation procedures.

Bad load balancing of the server processors was noticed in several experiments, and it was determined that for homogeneous client applications, this effect could be reduced through the use of narrow object groups. It is hypothesised that for heterogeneous client applications, the use of wide object groups may achieve better load balancing.

### 9.1.2 Object Migration

Three object migration schemes were tested. Both single object with prefetch and segment migration were found to waste space within the client processors' replica caches under some circumstances, but to yield benefits on other occasions.

Prefetching performs best with applications searching tree structures which have a low prefetch depth, branching factor, and memory access ratio. Its performance degrades gracefully as these conditions are eroded. Since the optimum prefetch depth for a structure is dependent upon these factors, it is believed that the prefetch depth should be an attribute of the class of objects forming the structure and be dynamically determined by the application.

Migration by segment was shown to yield marginal benefits when the spatial locality of storage matches the temporal locality of access, but to degrade performance appallingly when the two do not match. These results are for many objects per segment, and conflict with the findings Hosking and Moss report from their experiments with the Mneme object store[27, 15]. This is primarily because of architectural dissimilarities imposed by the nature of the underlying hardware, and indicates that common wisdom gained from experience with conventional object stores does not necessarily apply to stores running on multicomputer platforms.

### 9.1.3 Concurrency

The mutual exclusion provided by the use of binary semaphores on objects was found to be a workable but not easily programmable solution. The cost of maintaining coherence with this mechanism dominates applications using a fully shared data structure and seriously degrades efficiency. It was demonstrated that the facility to allow multiple readers during some computation phases could yield significant performance improvements.

## 9.2 Future Work

This project has given many valuable insights into issues concomitant with the efficiency of a multicomputer object store. However, some of the issues initially outlined do remain untouched, and others have arisen during the course of this work. Some of these are outlined below, along with possible lines of investigation.

It was demonstrated that allowing the application program to dynamically change the space recovery strategy gave improved performance. A preferable option for the application programmer would be for the LOM of the distributed object server to automatically switch between strategies as appropriate. This would require the LOM being capable of identifying which and when new strategies should be invoked. However, it would alleviate the programmer of the need to gain and then apply an understanding of the different space recovery strategies available. Perhaps the easiest change in access patterns to identify and accommodate would be to distinguish between object creation and general computation phases of an application.

The benefits of prefetching are presently restricted by several factors. It is believed that the region of benefit can be significantly expanded by conservative identification of objects likely to form part of a client's working set in the near future. This may be achieved by the identification of simple access patterns (full traversal as opposed to search for instance), or by utilising information such as object access counts, or common cofetch patterns. Additionally, automatic variation of prefetch depth dependent upon such factors as server load, free space in the client, and the data structure's branching factor could form an interesting investigation.

The efficiency implications of the binary semaphore model make evaluation of different coherence maintenance mechanisms a necessary avenue of further research. It is hoped that investigation of different locking mechanisms, optimistic scheduling, or loose coherence will provide a model which is both easier to program and more efficient.

The addition of a stable backing store to the system is the next obvious step in increasing functionality. This would facilitate experimentation with a broader range of data hungry applications and enable a true evaluation of the benefit of the object store paradigm in utilising multicomputers to service these applications.

# Glossary

**API:** Application Programming Interface: a software layer on the client side of the distributed object server, the layer with which user applications interact

**base server:** one of a set of servers responsible for a group, it is responsible for coordinating the creation of objects within the group

**binary semaphore:** a data type used for controlling concurrency. A binary semaphore is initialised to the value one. Subsequently, the operations Wait and Signal may be applied to it. Signal changes the semaphores value from zero to one, and Wait blocks the calling process until the semaphore's value is one and then changes it to zero. If multiple processes are blocked by Waits on the same semaphore, they are placed in a FIFO queue. It is an error for a process to Signal on a semaphore whose value is one, or to execute two Waits on a semaphore without an intervening Signal.

**CELLOS:** the operating system of the AP1000, mainly provides message passing routines

**deadlock:** a situation in which each possible activity is blocked, waiting on some other blocked activity

**GOM:** Global Object Manager: 1. a software layer on the server side of the distributed object server, which receives and attends requests from clients and other servers; 2. the collection of all GOM (1) instances

**LOM:** Local Object Manager: a software layer on the client side of the distributed object server, which manages the replica cache of the client and interacts with the *GOM* as necessary to create and fetch objects, and maintain coherence for concurrently accessed objects

**LRU:** Least Recently Used

**main server:** a single server, which is charged with performing GOM tasks that are not efficiently distributable

**master copy:** the instance of an object residing in the *store* (1)

**memory access ratio:** the ratio of the total size of all objects accessed by a client application, to the size of its replica cache

**MIMD:** Multiple Instruction, Multiple Data: a class of parallel machine capable of concurrently executing multiple instruction streams on multiple data sets

**MPI:** Message Passing Interface: a standard interface for message passing libraries, proposed by the Message Passing Interface Forum with the aim of increasing the portability of message passing programs, see [12]

**MPP:** Massively Parallel Processing

**MRU:** Most Recently Used

**object migration:** 1. the transport of an object or a replica from one processor to either another processor or a stable storage; 2. the transport of a replica from a server processor to a client processor, as opposed to *space recovery* (2)

**object reference:** a data type accessible to application programs, instances of which identify an object in the *store* by either containing an *OID* or a pointer to a *ROT* entry

**OID:** Object Identifier: uniquely identifies an object stored within the shared distributed heap managed by the distributed object server

**prefetch:** a method used by the GOM (2) to preemptively supply client processors with objects, based upon the references contained within fetched objects

**PVM:** Parallel Virtual Machine: a message passing library which runs on a variety of platforms and can be used to construct a virtual machine from a network of heterogeneous machines, see [13, 19]

**reference:** see *object reference*

**replica:** an instance of an object other than the *master copy*

**ROT:** Resident Object Table: a table maintained by the *LOM* of each client process which contains an entry for at least each resident replica; all object accesses pass through this table

**segment:** the unit of storage in which objects are stored within the servers, contains a data area and a segment directory

**SL:** Storage Layer: 1. a software layer on the server side of the distributed object server, which manages the segments stored in the server's memory; 2. the collection of all SL instances

**space recovery:** 1. the removal of unusable or unwanted objects from an object store; 2. the transport of a replica from a client processor to a server processor, as opposed to *object migration* (2); 3. an algorithm to determine which replicas to remove from the replica cache

**store:** 1. the collection of segments managed by the *SL*; 2. an object store

**transient memory:** the random access memory attached to a processor via a bus